# Density Functional Theory-based Quantum Mechanics/Coarse-grained Molecular Mechanics: Theory and Implementation

*Alexander V. Mironenko and Gregory A. Voth*[*]

Department of Chemistry, Chicago Center for Theoretical Chemistry, James Franck Institute, and Institute for Biophysical Dynamics, The University of Chicago, Chicago, Illinois 60637

KEYWORDS: quantum mechanics/coarse-grained molecular mechanics, quantum mechanics/molecular mechanics, force matching, multiscale coarse-graining

ABSTRACT: Quantum mechanics/molecular mechanics (QM/MM) is a standard computational tool for describing chemical reactivity in systems with many degrees of freedom, including polymers, enzymes, and reacting molecules in complex solvents. However, QM/MM is less suitable for systems with complex MM dynamics due to associated long relaxation times, the high computational cost of QM energy evaluations, and expensive long-range electrostatics. Recently, a systematic coarse-graining of the MM part was proposed to overcome these QM/MM limitations in the form of the quantum mechanics/coarse-grained molecular mechanics (QM/CG-MM) approach. Herein, we recast QM/CG-MM in the density functional theory formalism and, by employing the force-matching variational principle, access the method performance for two model systems: QM $CCl_4$ in the MM $CCl_4$ liquid and the reaction of tert-butyl hypochlorite with the benzyl radical in the MM $CCl_4$ solvent. We find that DFT-QM/CG-MM accurately reproduces DFT-QM/MM radial distribution functions and 3-body correlations between QM and CG-MM subsystems. The free energy profile of the reaction is also described well, with an error < 1-2 kcal/mol. DFT-QM/CG-MM is a general, systematic, and computationally efficient approach to include chemical reactivity in coarse-grained molecular models.

## INTRODUCTION

Accurate simulation of chemical reactivity is crucial for the atomic-level understanding of natural processes involving chemical bond rearrangements (such as proton transfer in water[1,2] and biomolecules[3,4]) and for computer-aided material design[5]. To obtain reliable estimates of reaction energies and activation barriers in small to mid-size chemical systems, one typically begins with either the approximate density functional theory (DFT)[6–8] or more accurate wave function-based approaches (e.g., the coupled-cluster (CC) theory[9,10]) to determine relative energies of states in the configuration space. The potential energy field is then translated into reaction free energy profiles using either a combination of molecular dynamics[11,12] and enhanced sampling[13–18] methods or saddle point/minimum energy search techniques[19–22] together with models for temperature effects, such as the harmonic approximation within the transition state theory[23]. However, the first-principles-based approach is often too costly to be feasible for complex systems with bond breaking/formation coupled to the correlated motion of large quantities of atoms, due to unfavorable scaling of DFT and CC with the system size ($\sim N^3$ and $\sim N^7$, respectively). Notable examples of such systems include enzymes[24], stimuli-responsive polymers[25], and functionalized nanoparticles[26].

The computational cost of energy calculations in complex reactive systems can be drastically reduced by partitioning the system into reactive and non-reactive parts. Due to the "nearsightedness" of chemical interactions[27,28], it is often sufficient to use the full quantum-mechanical representation for only a relatively small number of atoms surrounding a chemical reaction site in a molecule. In contrast, the conformational behavior of the rest of the system with preserved interatomic connectivity can be accurately described using much less expensive classical molecular mechanical (MM) force fields[29–33]. Such a system partitioning has been first proposed by Warshel and Levitt in 1976[34] and forms the foundation of the method of quantum mechanics/molecular mechanics (QM/MM). QM/MM has gained widespread use in simulations of spectroscopic and excited-state properties in biological pigments and enzymes, as well as in modeling of the chemical reactivity in enzymes[35,36], with notable examples including hydroxylases, peroxidases[37], and nitrogenases[38].

Despite significant computational gains offered by QM/MM relative to full *ab initio* approaches, the methodology has its limitations. In the most commonly used electrostatic embedding scheme, evaluations of electrostatic interactions between QM and MM subsystems can easily take between 20% and 100% of the QM computational cost for plane-wave basis sets[39], adding significant overhead. Additionally, the ergodic sampling of the configuration space can become prohibitively expensive for the systems with many MM atoms, since the system relaxation time increases with the number of its degrees of freedom (DoFs)[40]. Thus, very long MD trajectories may be required. One example includes hydrolysis of adenosine triphosphate (ATP), attached to the filament of the protein actin, which participates in many vital functions of the cell, such as motility, division, and cytokinesis. The QM/MM MD study of the ATP hydrolysis kinetics, catalyzed by G-actin (actin monomer), required 5 million CPU hours to obtain 1.5 ns trajectories[41]. Since the ATP hydrolysis rate is also affected by the states of neighboring actin subunits[42], unraveling atomic-level details of the influence of the entire filament on the hydrolysis reaction is clearly beyond the capabilities of even QM/MM. Therefore, there is a need to develop QM/MM alternatives capable of tackling large reactive systems containing many MM atoms.

Coarse-grained (CG) methods are effective at reducing the complexity of all-atom (AA) MM-based representations of complex molecules, enabling long and large-scale simulations[43–46]. In the CG representation of a system, atoms map onto a fewer number of particles interacting through effective potentials. As a result of the CG mapping, less relevant high-frequency fluctuations of correlated atoms are integrated out, whereas physically significant low-frequency, correlated motions are retained in the form of CG sites or "beads" traversing the free energy surface (FES). The CG models provide considerable computational gains, as (1) fewer DoFs signify shorter relaxation times and shorter trajectories; (2) smoother FES in comparison with the underlying PES facilitates CG conformational space exploration; (3) long-range electrostatic and mid-range dispersion interactions that often dominate the cost of MM energy evaluations, are replaced by short-range renormalized interparticle potentials. Unlike QM/MM, the CG models, however, lack quantum mechanical details of a system, and thus, in general, are incapable of describing chemical reactivity, unless the topology-switching mechanism is included[47].

Herein, we report the development and implementation of a DFT-based mixed-resolution quantum mechanics/coarse-grained molecular mechanics (QM/CG-MM) methodology that incorporates quantum mechanical information into CG models in a rigorous manner. By coarse-graining the non-reactive part and retaining the AA QM description of the reactive part of a system, the method aims to alleviate QM/MM shortcomings in large biomolecular systems by leveraging computational gains of the CG representations. The rigorous theoretical framework for the QM/CG-MM method has been put forward by Sinitskiy and Voth[48], using a wave mechanics and perturbation theory formalism. In this work, we reformulate the theory in the density functional theory language and discuss implementation and performance of DFT-QM/CG-MM using two model systems: $CCl_4$ liquid and a reaction of tert-butyl hypochlorite with benzyl radical in the $CCl_4$ solvent.

## THEORY

**DFT-QM/MM.** As a starting point, we begin with the additive QM/MM total energy expression in the electrostatic embedding scheme. Appendix presents its rigorous derivation from the Kohn-Sham (KS) DFT, which, to the best of our knowledge, has not been reported previously. The QM/MM total energy in the additive scheme is

$$E_{QM/MM}(\mathbf{R}^N) = E_{QM}(\mathbf{R}^N; \rho_{QM}) + E_{MM \leftrightarrow MM}(\mathbf{R}^{N_{MM}}) + E_{QM \leftrightarrow MM}(\mathbf{R}^N), \quad (1)$$

where $E_{QM}$ and $\rho_{QM}$ are the energy and the electron density of the QM region; $E_{MM \leftrightarrow MM}$ and $\mathbf{R}^{N_{MM}}$ are the energy and atomic coordinates of the MM region consisting of $N_{MM}$ atoms; $\mathbf{R}^N = \mathbf{R}^{N_{QM}} \times \mathbf{R}^{N_{MM}}$, where $\mathbf{R}^{N_{QM}}$ are the nuclear coordinates of $N_{QM}$ atoms in the QM region; and $E_{QM \leftrightarrow MM}$ is the interaction energy of QM and MM subsystems. For the DFT QM energy we assume semilocal approximation of XC effects[8] with added perturbative dispersion corrections[49,50], yielding

$$E_{QM}(\mathbf{R}^{N_{QM}}; \rho_{QM}) = \sum_i f_i^{QM} \varepsilon_i^{QM} -$$

$$\int \rho_{QM}(\mathbf{r}) \left( \frac{1}{2} \phi[\mathbf{r}; \rho_{QM}] + \mu_{xc,sl}(\rho_{QM}(\mathbf{r})) + V_{es}^{MM}(\mathbf{r}; \mathbf{R}^{N_{MM}} | \mathbf{q}_{MM}) \right) d\mathbf{r} \quad (2)$$

$$+ E_{xc,sl}[\rho_{QM}] + E_{vdW}(\mathbf{R}^{N_{QM}}) + E_{NN}(\mathbf{R}^{N_{QM}})$$

$$+ \int \rho_{QM}(\mathbf{r}) V_{es}^{MM}(\mathbf{r}; \mathbf{R}^{N_{MM}} | \mathbf{q}_{MM}) d\mathbf{r},$$

where $\varepsilon_i^{QM}$ and $f_i^{QM}$ are the energies and occupancies of KS one-electron states, $\phi$ is the Hartree potential, $E_{xc,sl}$ is the semilocal XC energy, $\mu_{xc,sl} = \partial E_{xc,sl}/\partial \rho$ is the semilocal XC potential, $E_{NN}$ and $E_{vdW}$ are the internuclear electrostatic repulsion energy and the energy of van der Waals interactions in the QM region, $V_{es}^{MM} = \sum_{i \in MM} q_{MM,i}/|\mathbf{r} - \mathbf{R}_i|$ is the electrostatic potential of $q_{MM,i} \in \mathbf{q}_{MM}$ MM point charges with $\mathbf{R_i} \in \mathbf{R}^{N_{MM}}$. The last term represents the double counting correction, as the QM-MM electrostatic interaction is already included in $E_{QM \leftrightarrow MM}$. Energies $\varepsilon_i^{QM}$ and their corresponding states $\psi_i^{QM}$ are the solutions to the KS equations (atomic units used),

$$\left(-\frac{1}{2}\nabla^2 + v_N(\mathbf{r}; \mathbf{R}^{N_{QM}}) + \phi[\mathbf{r}; \rho_{QM}] + \mu_{xc,sl}\left(\rho_{QM}(\mathbf{r})\right) \right. \\ \left. + V_{es}^{MM}(\mathbf{r}; \mathbf{R}^{N_{MM}}|\mathbf{q}_{MM})\right)\psi_i^{QM}(\mathbf{r}) = \varepsilon_i^{QM}\psi_i^{QM}(\mathbf{r}), \quad (3)$$

where $v_N$ is the external (nuclear) potential. The QM electron density is defined as $\rho_{QM}(\mathbf{r}) = \sum_i f_i |\psi_i^{QM}(\mathbf{r})|^2$, and the states $|\psi_i\rangle$ are assumed to be expanded in the unspecified atomic basis set. In eq. (1), the QM/MM interaction energy is $E_{QM \leftrightarrow MM}(\mathbf{R}^N) = E_{vdW,QM \leftrightarrow MM}(\mathbf{R}^N) + E_{ortho,QM \leftrightarrow MM}(\mathbf{R}^N) + E_{es,QM \leftrightarrow MM}(\mathbf{R}^N)$, where $E_{vdW,QM \leftrightarrow MM}$ and $E_{ortho,QM \leftrightarrow MM}$ are the van der Waals and orthogonalization (Pauli repulsion) contributions to the interaction energy, and the electrostatic contribution $E_{es,QM \leftrightarrow MM}$ equals the sum of point charge↔nuclear and point charge↔electronic contributions $E_{NN,QM \leftrightarrow MM}(\mathbf{R}^N|\mathbf{q}_{MM}) + \int \rho_{QM}(\mathbf{r}) V_{es}^{MM}(\mathbf{r}; \mathbf{R}^{N_{MM}}|\mathbf{q}_{MM})d\mathbf{r}$.

**DFT-QM/CG-MM.** In the CG representation of the all-atom (AA) MM subsystem, it is natural to express the total DFT-QM/CG-MM energy of the system as a sum of terms, in analogy with eq. (1):

$$E_{QM/CG-MM}(\mathbf{R}^N) = E_{QM}(\mathbf{R}^N; \rho_{QM}) + E_{CG \leftrightarrow CG}(\mathbf{R}^{N_{CG}}|\mathbf{K}_{CG}) + E_{QM \leftrightarrow CG}(\mathbf{R}^N|\mathbf{L}_{CG}) \quad (4)$$

where $\mathbf{R}^{N_{CG}}$ are the coordinates of the $N_{CG}$ CG particles; $\mathbf{R}^N = \mathbf{R}^{N_{QM}} \times \mathbf{R}^{N_{CG}}$, and $E_{CG \leftrightarrow CG}$ and $E_{QM \leftrightarrow CG}$ are yet unspecified interparticle interaction energies within the CG subsystem and between CG and QM parts, respectively. The symbols $\mathbf{K}_{CG}$ and $\mathbf{L}_{CG}$ denote groups of parameters defining the corresponding interactions. The DFT-QM/CG-MM equivalent of eq. (3) takes the form

$$\left(-\frac{1}{2}\nabla^2 + v_N(\mathbf{r}; \mathbf{R}^{N_{QM}}) + \phi[\rho_{QM}(\mathbf{r})] + \mu_{xc,sl}[\rho_{QM}(\mathbf{r})] \right. \\ \left. + V_{es}^{CG}(\mathbf{r}; \mathbf{R}^{N_{CG}}|\mathbf{M}_{CG})\right)\psi_i^{QM}(\mathbf{r}) = \varepsilon_i^{QM}\psi_i^{QM}(\mathbf{r}), \quad (5)$$

where $V_{es}^{CG}(\mathbf{r}; \mathbf{R}^{N_{CG}}|\mathbf{M}_{CG})$ is the yet unspecified CG representation of the electrostatic potential emitted by the coarse-grained MM subsystem and determined by parameters $\mathbf{M}_{CG}$.

In the following, we focus on the determination of CG parameter vectors $\mathbf{K}_{CG}$, $\mathbf{L}_{CG}$, and $\mathbf{M}_{CG}$. The critical requirement of any rigorous CG model is that it must obey the consistency condition [51,52], which implies the equality between probability densities in the configuration space of CG and CG-mapped AA representations. Mapping the AA trajectories onto CG beads is usually accomplished using the linear coordinate transformation $\mathbf{R}_{CG,i} = \sum_{j \in i} c_j \mathbf{R}_{MM,j}$, where $\mathbf{R}_{MM,j}$ and $\mathbf{R}_{CG,i}$ are particle coordinates in the AA and CG representation, respectively. The weighing coefficients $c_j$ are most frequently correspond to the center-of-mass operator. It can be shown[53] that the consistency condition is equivalent to the minimization of a proper variational functional, with two notable examples being the mean squared difference between CG and AA forces (the force-matching method)[54] and the relative entropy[55]. Herein, we adopt the force-matching methodology to the DFT-QM/CG-MM framework and use it to determine $\mathbf{K}_{CG}$, $\mathbf{L}_{CG}$, and $\mathbf{M}_{CG}$ parameters from all-atom QM/MM trajectories.

In the most general case, the variational functional for the force matching is given by

$$\chi^2[V_{CG}] = \frac{1}{3N}\left\langle \sum_I^{N_{CG}} |f_I(\mathbf{R}^{N_{CG}}) - F_I(\mathbf{R}^{N_{CG}}|V_{CG})|^2 \right\rangle, \quad (6)$$

where $V_{CG}$ is the CG potential to be determined, $f_I(\mathbf{R}^{N_{CG}})$ are the AA forces projected onto CG particles, $F_I$ are the forces due to $V_{CG}$, and the averaging is performed over an AA MD trajectory. To adapt eq. (6) to DFT-QM/CG-MM, we first recognize that both DFT-QM/MM and DFT-QM/CG-MM forces can be represented as sums of QM and non-QM contributions. For the DFT-QM/CG-MM case, we have

$$F_{QM/CG-MM}(\mathbf{R}^N) = F_{QM}(\mathbf{R}^N; \rho_{QM}|\mathbf{M}_{CG}) + F_{non-QM}(\mathbf{R}^N|\mathbf{K}_{CG}, \mathbf{L}_{CG}), \quad (7)$$

where $F_{QM/CG-MM} = -\nabla E_{QM/CG-MM}$, $F_{QM} = -\nabla E_{QM}$, and $F_{non-QM} = -\nabla E_{CG-CG} - \nabla E_{QM-CG}$. Next, we introduce two force-matching conditions, held separately for $F_{QM}$ and $F_{non-QM}$:

$$\chi^2(\mathbf{M}_{CG}) = \frac{1}{3N_{QM}}\left\langle \sum_I^{N_{QM}} |f_{QM,I}(\mathbf{R}^{N_{QM}}|\mathbf{q}_{MM}) - F_{QM,I}(\mathbf{R}^{N_{QM}}|\mathbf{M}_{CG})|^2 \right\rangle, \quad (8)$$

and

$$\chi^2(\mathbf{K}_{CG}, \mathbf{L}_{CG}) = \frac{1}{3N}\left\langle \sum_I^{N} |f_{QM/MM,I}(\mathbf{R}^N) - f_{QM,I}(\mathbf{R}^{N_{QM}}|\mathbf{q}_{MM}) - F_{non-QM,I}(\mathbf{R}^N|\mathbf{K}_{CG}, \mathbf{L}_{CG})|^2 \right\rangle, \quad (9)$$

where $\mathbf{f}_{QM,I}$ and $\mathbf{f}_{QM/MM,I}$ are the QM and total forces in the reference DFT-QM/MM calculation. In the ideal case of a perfect representation of AA forces by the CG model, $\chi^2(\mathbf{M}_{CG}) = \chi^2(\mathbf{K}_{CG}, \mathbf{L}_{CG}) = 0$, and thus in eq. (6) $\chi^2[V_{CG}] = 0$, in accordance with the consistency condition.

Since $F_{QM,I}$ (the QM force in the DFT-QM/CG-MM method) is a functional of $V_{es}^{CG}$ (see eq. (5)), it is evident that for the ideal case of $\chi^2(\mathbf{M}_{CG}) = 0$ in eq. (8) we must have the equality $V_{es}^{CG}(\mathbf{r}; \mathbf{R}^{N_{CG}}|\mathbf{M}_{CG}) = V_{es}^{MM}(\mathbf{r}; \mathbf{R}^{N_{MM}}|\mathbf{q}_{MM})$, i.e., the CG electrostatic potential should exactly

reproduce the MM point charge potential. Consequently, eq. (8) is equivalent to the following "potential matching" condition:

$$\chi^2(\mathbf{M}_{CG}) = \frac{1}{3N_{QM}} \left\langle \int d\mathbf{r} \, |V_{es}^{MM}(\mathbf{r}; \mathbf{R}^{N_{MM}}| \mathbf{q}_{MM}) - V_{es}^{CG}(\mathbf{r}; \mathbf{R}^{N_{CG}}|\mathbf{M}_{CG})|^2 \right\rangle, \quad (10)$$

By minimizing eq. (10) and eq. (9), one can get CG parameters $\mathbf{K}_{CG}, \mathbf{L}_{CG}, \mathbf{M}_{CG}$ for the chosen $V_{es}^{CG}$ and $F_{non-QM,I}$ functional forms.

For the DFT-QM/CG-MM method implementation described herewith, we make one additional approximation. We take advantage of the fact that the QM energy functional is variational with respect to the KS potential[56], and thus first-order errors in the potential lead to second-order errors in the energy. As free energy errors are proportional to potential energy errors to the first order from perturbation theory, we conclude that the approximate $V_{es}^{CG}$ should have a minor influence on thermodynamics. Consequently, we invoke the monopole approximation by defining CG particle charges $Q_I = \sum_i q_{MM,i}$ with $i \in I$ and the CG electrostatic potential as

$$V_{es}^{CG} = \sum_I \frac{Q_I}{|r - r_I|}. \quad (11)$$

## IMPLEMENTATION

**Software.** We implemented the method on the basis of the CP2K[57] software [48], modified in-house for DFT-QM/CG-MM. CP2K delivers the most general, fully periodic implementation of QM/MM, with a highly efficient Gaussian expansion of the electrostatic potential (GEEP) electrostatic embedding scheme[39]. Additionally, CP2K employs the mixed-basis gaussian/plane-wave (GPW) scheme[58], in which KS states are expanded in the local gaussian (GTO) basis set, whereas the electron density is expanded in an auxiliary basis set consisting of plane waves. Accordingly, all one-electron integrals are evaluated analytically, while the two-electron four-center integrals, often dominating the cost of energy evaluations in purely GTO methods[59], are eliminated and replaced by efficient Fast Fourier Transform-based techniques. High efficiency, robustness, and widespread use of CP2K were the main factors in selecting the software for DFT-QM/CG-MM implementation.

**Algorithm.** A key ingredient of the DFT-QM/CG-MM method are the CG potentials describing effective interactions between CG beads $[E_{CG \leftrightarrow CG}(\mathbf{R}^{N_{CG}}|\mathbf{K}_{CG})]$ and the QM and CG subsystems $[E_{QM \leftrightarrow CG}(\mathbf{R}^N|\mathbf{L}_{CG})]$. Figure 1a depicts the algorithm for generating CG potentials. We begin with an all-atom DFT-QM/MM simulation of a system of interest and output atomic trajectories and *non-QM* forces $f_{QM/MM,I}(\mathbf{R}^N) - f_{QM,I}(\mathbf{R}^{N_{QM}}|\mathbf{q}_{MM})$ [see eq. (9)] to disk. In the next step, we map AA trajectories and non-QM forces onto CG trajectories and forces using the center-of-mass CG mapping operator as follows (Figure 1b):

$$R_{CG,I} = \frac{\sum_{j \in I} m_j R_{AA,j}}{\sum_{j \in I} m_j}, \quad j \in MM \tag{12}$$

$$R_{CG,I} = R_{AA,j}, \quad j \in QM$$

where $m_j$ is the mass of atom $j$. To avoid computing different potentials for chemically similar atoms, we divide QM atoms into $N_{QM,g}$ groups according to the force field atom type assignments (OPLS-AA[32] in this work). Consequently, the number of pairwise CG interactions to be determined equals $N_{QM,g} \times N_{CG,g} + N_{CG,g}(N_{CG,g} + 1)/2$, where $N_{CG,g}$ is the number of CG bead types. We then use the output of the mapping procedure as an input to the Multiscale Coarse-Graining (MS-CG) code[60], developed in our group, which implements eq. (6). The code produces numerically tabulated CG potentials to be used as an input to DFT-QM/CG-MM calculations in CP2K. For setting up CP2K calculations, we integrated several tools, including MOLTEMPLATE[61], PACKMOL[62], VMD[63], and the LAMMPS-to-CP2K force field converter, developed in-house.

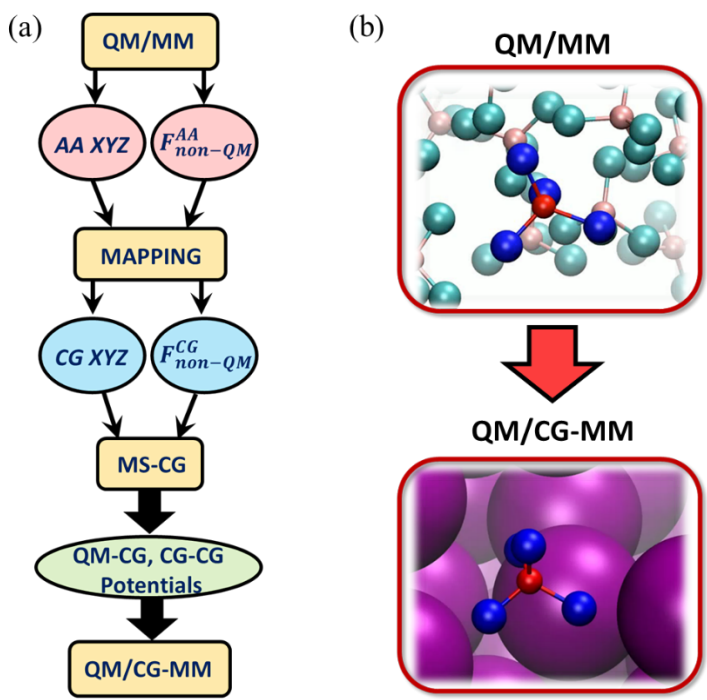

Figure 1. (a) Information workflow for parameterization of QM/CG-MM. (b) Transitioning from the QM/MM to the QM/CG-MM representation using "QM CCl$_4$ in MM CCl$_4$" as an example. Dark red (C) and blue (Cl) atoms belong to the QM region. Light red (C) and cyan (Cl) atoms belong to the MM region. Purple particles are the CG representation of MM CCl$_4$ molecules.

The "bottom-up" approach of generating CG potentials from AA trajectories, shown in Figure 1a, requires some degree of QM/MM calculation for the method parameterization, which may be expensive. To address this possible shortcoming, below we will demonstrate the method's insensitivity to (1) the choice of the QM method, (2) the fine details of CG potentials. Consequently, we will show that relatively short, semiempirical QM/MM trajectories may be sufficient to obtain CG potentials, accurate enough to be used in higher-level DFT-QM/CG-MM calculations, at least for the model systems studied herein.

**Tabulated vs. analytical potentials.** The FIST molecular mechanics engine of CP2K only allows for analytical functional forms of the custom potentials, not the tabulated potentials, such as those generated by the MS-CG code. As analytical potentials, such as Lennard-Jones, are frequently utilized in some more "top-down" coarse-graining approaches[64], we decided to assess their adequacy for reproducing the underlying AA probability distributions. To this end, we constructed a fully classical CG model of the $CCl_4$ liquid with one CG particle per molecule and first obtained tabulated CG potentials from an AA classical simulation. Then, MS-CG potentials were approximated by the form $E = wE_{asc} + (1-w)E_{desc}$ with ascending $E_{asc}$ and descending $E_{desc}$ energy fits of the general form $A_i/r^{n_i} + B_i/r^{m_i}$ and the switching function $w = 1/(1 + e^{-2k(r-r_0)})$. $A_i$ and $B_i$ were constrained to match the function value and its derivative at the energy minimum $r_0$, to ensure that the dominant lowest-energy region of the potential is reproduced most accurately. Both potentials before and after analytical approximation are shown in Figure 2a.

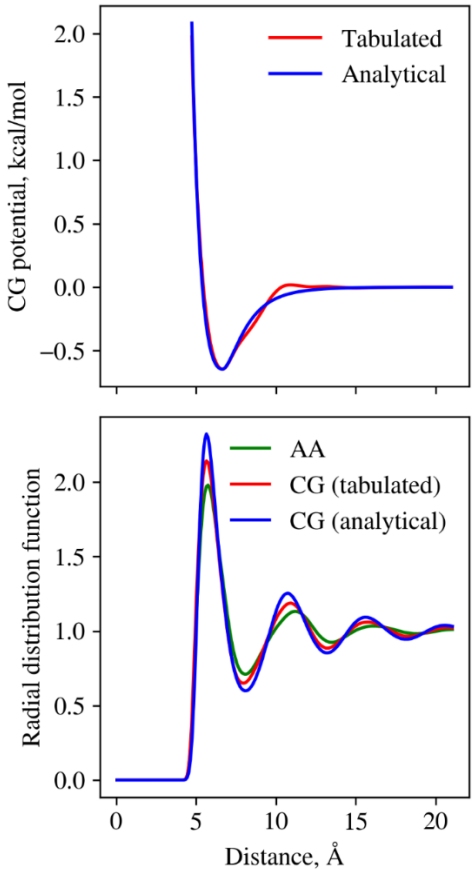

Figure 2. (a) Tabulated (red) and analytical (blue) coarse-grained potentials acting between $CCl_4$ CG beads. (b) Radial distribution functions (RDF) between $CCl_4$ CG beads generated in classical CG MD simulations using the tabulated (red) or analytical (blue) potentials. Center-of-mass RDF from the all-atom simulation is shown for comparison (green).

Figure 2b displays radial distribution functions $g(r)$ between the molecular center of masses (COM), as obtained from the AA (blue line), tabulated CG (red line), and analytical CG (green line) simulations. Although coarse-graining introduces a minor error in spatial correlations relative to the AA system, which is common[51], the analytical approximation increases this error somewhat. This observation prompted us to not utilize analytical CG potentials in DFT-QM/CG-MM, as it was critical to bracket the sources errors when testing the method, so we instead implemented tabulated custom potentials in CP2K, using the linear spline interpolation. To illustrate the accuracy of the interpolation scheme, in Figure S1 we compare $g(r)$, computed using identical tabulated potentials in LAMMPS[65] and in the modified CP2K, respectively. The radial distributions are indistinguishable. It should be noted, however, that many users prefer analytical potentials and some codes that might be modified to perform QM/CG-MM simulations may require them, so we retained these illustrative results here.

**Choice of model systems.** To assess the performance of the DFT-QM/CG-MM method, we considered two simple model systems. The first one is the "$CCl_4$-in-$CCl_4$" liquid at 300 K, in which one $CCl_4$ molecule is treated quantum mechanically, whereas the rest (499 molecules) are treated classically either at the AA resolution (as in QM/MM) or at the CG resolution with one CG bead per molecule (in QM/CG-MM). The second system consists of the tert-butyl hypochlorite (TBHC) reacting with the benzyl radical in the $CCl_4$ solvent at 313.15 K. This reaction is known to be the rate-limiting propagation step in the free-radical chain chlorination of toluene[66]. Both reactants are treated quantum mechanically (29 atoms total), and 500 CCl$_4$ solvent molecules are included either in the MM (in QM/MM) or the CG region (in QM/CG-MM). The use of the non-polar solvent provides considerable simplifications, as no anisotropic nonbonded interactions arise (such as those present in liquid water), and all CG beads have zero charges in the monopole approximation employed here.

## COMPUTATIONAL DETAILS

The QM/MM and QM/CG-MM calculations were carried out in CP2K 6.1. The CCl$_4$ liquid model (Figure 1b) consisted of a single CCl$_4$ molecule treated quantum-mechanically and 499 molecules or CG beads described classically. The optimal edge length of a cubic simulation cell (43.18 Å) was determined by means of 1 ns-long classical MD simulations in the grand canonical ensemble with the Nose-Hoover thermostat[67,68]/barostat[69], as implemented in LAMMPS[65]. The reactive model system consisted of a benzyl radical and a tert-butoxy hypochlorite (TBHC) species, described quantum-mechanically, and 500 classical CCl$_4$ molecules or CG beads. The same simulation box size was used.

The QM region was modeled using the Kohn-Sham DFT formalism and the mixed-basis GPW computational scheme[70]. The valence KS states were expanded in the gaussian-type basis set of double-zeta quality with polarization functions (DZVP)[71]. Certain calculations used the single-zeta basis set (SZV). Effects of core electrons were approximated by analytical Goedecker-Teter-Hutter pseudopotentials[72]. The electron density was expanded in the auxiliary plane wave basis set. Both gaussian orbitals and electron density were mapped onto four commensurate multigrids of varying coarseness in the real space with the cutoff of the finest grid level set to 150 Ry. The cutoff for the gaussian reference grid was set to 40 Ry. We employed the orbital transformation (OT) method[73] to solve the KS energy minimization problem. For the QM region of the CCl$_4$ liquid, we used the minimizer based on the Broyden mixing approximation of the inverse Hessian. We also selected the diagonalization-based state-selective preconditioner with the chosen energy gap of 0.1 hartrees. For the spin-unrestricted computation of the radical reaction energetics (multiplicity=2), we employed the conjugate gradient minimizer with the same preconditioner, iterative algorithm, and the Lowdin orthogonalization method. Both inner and outer SCF loops were used with the target accuracy of $10^{-5}$. The XC effects were approximated by the PBE functional[8].

The MM region was modeled using the OPLS-AA[32] force field, specifically developed for liquids. Force field parameters were taken from the oplsaa.lt file, downloaded from www.moltemplate.org. MM neighbor lists were generated using the 2 Å-thick Verlet skin. Electrostatic interactions were computed using the smooth particle mesh method[74] with the alpha parameter set to 0.3 Å, 43 grid points, and the order of beta-Euler splines equal to 6. The cutoff for nonbonded interactions was set to 10 Å.

In QM/MM and QM/CG-MM calculations, we employed the electrostatic coupling with the expansion of the electrostatic potential in terms of 10 gaussian functions (GEEP)[39]. The size of the QM box was chosen to be 8×8×8 Å for CCl$_4$ (DZVP basis and DFTB), 9×9×9 Å (SZV basis) and 14×14×14 Å for the reactive system. Classical point charges were smoothed by a gaussian function using standard tabulated covalent radii[75]. GMAX was set to 0.5. QM periodic images were decoupled using the multipole scheme[76] with the real space cutoff of 10 Å. We employed reflective walls to confine the QM region inside the QM box. The effect of box size on potentials was found to be statistically insignificant. The QM box was recentered at every step in units of grid spacing.

A selected number of QM/MM and QM/CG-MM calculations employed the approximate self-consistent charge density functional tight binding (SCC-DFTB) method[77] to reduce the computational cost. The 3ob-3-1 file containing method's parameters was taken from the www.dftb.org website. We used the Coulomb electrostatic embedding and set the QM box dimensions to be equal to those of the simulation box. The QM walls were turned off.

Unbiased MD simulations were carried out in the canonical ensemble at 300 K ($CCl_4$) and 313.15 K (reactive system). The temperature was controlled using the Nose-Hoover thermostat chain[67,68] with default parameters. Verlet[12] integration was carried out with a 1 fs time step. Trajectories and forces were written to disk every 50-100 fs.

Biased MD simulations were carried out using the umbrella sampling technique, as implemented in the PLUMED2[78] software package, integrated with CP2K. The CV range was [-4.0; 4.0] Å; umbrella windows were placed every 0.25 Å. We employed the harmonic biasing potential with the force constant of 80 kcal/(mol Å); for QM/CG-MM at CV=2.0 Å the harmonic constant was set to 140 kcal/(mol Å). Optimal force constants were found by trial-and-error to ensure that the CV probability distributions are centered at the respective windows. Biased CV probability distributions were converted to the free energy profile by means of the weighted histogram analysis method (WHAM)[16], as implemented by Grossfield[79]. The 95% confidence intervals were found by performing three independent umbrella sampling calculations that used different initial structures. The duration of umbrella sampling trajectories in each window was 36-45 ps for QM/MM and 40-50 ps for QM/CG-MM. The first 8 ps and 4 ps of trajectories have been discarded for QM/MM and QM/CG-MM, respectively.

To generate statistically meaningful results, we adopted the following sequence of operations. For the QM/MM (as well as QM/CG-MM) simulations of the $CCl_4$ liquid, we carried out the geometry optimization of the MM region for 500 fs while keeping QM atoms frozen, followed by 1 ns simulations in the canonical ensemble using either DFT/SZV, DFT/DZVP, or DFTB methods. The 10 ps-long equilibration trajectory was discarded. Non-QM forces and atomic coordinates were written to disk every 50 fs and were subsequently projected onto CG particles. The force-matching MS-CG computations were carried out using the order-3 b-spline pairwise basis set with the resolution of 0.2 Å. The tabulated potential was recorded every 0.01 Å.

For the QM/MM unbiased simulations of the reactive system, we similarly performed 500 fs-long optimization of the MM region, followed by 1 ns unbiased simulations. Initial geometries for the umbrella sampling were prepared by steered MD followed by the constrained energy minimization.

## RESULTS AND DISCUSSION

**The "CCl$_4$-in-CCl$_4$" system.** *CG potentials*. In Figure 3 we report CG potentials describing interactions between QM atoms and CG beads in the CCl$_4$ liquid, as obtained by force-matching non-QM forces from 1 ns-long QM/MM trajectories for three levels of the QM theory (DFT/SZV, DFT/DZVP, and DFTB, where SZV and DZVP are single-zeta/double-zeta basis sets used). The CG-CG potential exhibits a minimum at ~7 Å with a depth of ~ – 0.5 kcal/mol, which corresponds to weak, mainly van der Waals interactions between CG particles. Its shape is nearly identical to its classical analog parameterized on the MM CCl$_4$ model (not shown), as the effect of a single QM molecule on interactions among 499 MM molecules is very small. For the same reason, the CG-CG potential is weakly sensitive to the QM method chosen.

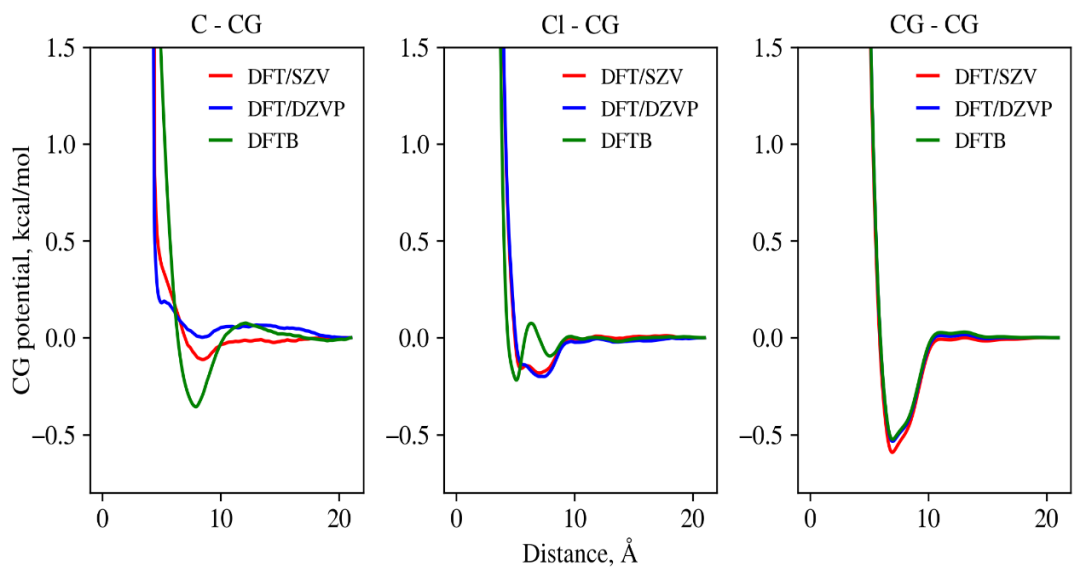

Figure 3. Non-QM CG potentials for the CCl$_4$ liquid, obtained by force-matching of ~ 1 ns QM/MM trajectories. Standard deviations at energy minima for C-CG, Cl-CG, and CG-CG interactions are 0.11, 0.04, and 0.12 kcal/mol (DFT/SZV); 0.01, 0.02, and 0.02 kcal/mol (DFT/DZVP); and 0.07, 0.02, and 0.03 kcal/mol (DFTB), respectively. Standard deviations were computed by splitting QM/MM trajectories into 3 pieces of equal length and using the force matching procedure on each of them.

Unlike CG-CG, the C-CG and Cl-CG potentials exhibit different behavior. The DFT-level Cl-CG potential has a broad minimum with the depth of -0.2 kcal/mol at ~7-7.5 Å. Curiously, we find that energy minima positions for both Cl-CG and CG-CG potentials are similar, likely due to the symmetry of the CCl$_4$ molecule and homogeneous conditions in the liquid. This can be explained as follows: if we consider the rotational+vibrational motion of a single CCl$_4$ molecule around a COM, the probability of finding a Cl atom will take a form of a spherically symmetric shell. Consequently, the CG bead interacting with the CCl$_4$ molecule will on average "see" two equivalent Cl positions on a line connecting CCl$_4$ and CG COMs, equidistant from the CCl$_4$ COM. Due to symmetry, the effect of the Cl atom "cloud" will be equivalent to that of an "effective Cl" atom placed near the CCl$_4$ COM, explaining the similarity of minima positions of Cl-CG and CG-CG potentials.

Another feature of SZV and DZVP Cl-CG potentials is that they are nearly identical, reflecting a weak sensitivity of underlying QM↔MM electrostatic interactions to the basis set. This follows from the Gauss law, as well as from small QM polarization effects due to homogeneous conditions in the CCl$_4$ MM liquid. The DFTB Cl-CG potential, however, has a different shape with two shallow minima located near 5 and 8 Å. Similar differences arise for the C-CG potential, where the DFTB-derived potential exhibits a single deep minimum near 7.9 Å, while DFT-based C-CG potentials are much shallower (Figure 3). We attribute potential shape variations between DFT and DFTB to different formulations of QM↔MM electrostatic interaction (ES) energies in both methods. In DFT, the ES energy is computed as the integral over the electron density and Gaussian-smoothed MM point charges, whereas in DFTB, the ES energy is approximated by the ES interaction between MM and QM-derived Mulliken point charges, and no electron density is explicitly constructed[80].

A notable feature of C-CG and Cl-CG potentials is substantial uncertainty in their numerical values, as manifested by large standard deviations relative to minima depths in Figure 3, with even larger values at higher energies (not shown). It partially arises due to limited statistics: although there are $N(N-1)/2$ CG-CG pairs in the simulation box with $N = 499$, there are only $N$ C-CG and $4N$ Cl-CG pairs. Also, C and Cl particles are more mobile than CG due to their lower mass, so that on a given interparticle distance interval $[R, R + \Delta R]$ there is a fewer number of particle pairs sampled from the MD trajectory. A key question here is how sensitive the observables are to the potential shape variations in Figure 3. This is discussed next.

*Force distributions.* Figure 4 depicts distributions of QM and non-QM forces on C and Cl atoms in the QM region, obtained from DFT-QM/MM and DFT-QM/CG-MM. The QM force distributions (Figure 4a and b) demonstrate near-perfect overlap, signifying the accuracy of the monopole approximation of the CG electrostatic potential (eq. (11)). Non-QM force distributions (Figure 4c and d) show considerable overlap but deviate more significantly in all regions. Notably, the DFT-QM/CG-MM force distribution on Cl atoms is more localized and lacks the high-force tail. Although it is difficult to pinpoint the exact cause of such localization, we hypothesize that it is due to the lack of strong long-range electrostatic interactions between QM and CG subsystems, which may become important when the CG particle passes in the vicinity of the QM region. Deviations can also be due to the large uncertainty and slow convergence of CG potentials, an issue discussed earlier. In order to see whether force deviations are consequential to the reproducibility of QM/MM observables by the CG model, we turn to the comparison of the radial distribution functions (RDF).

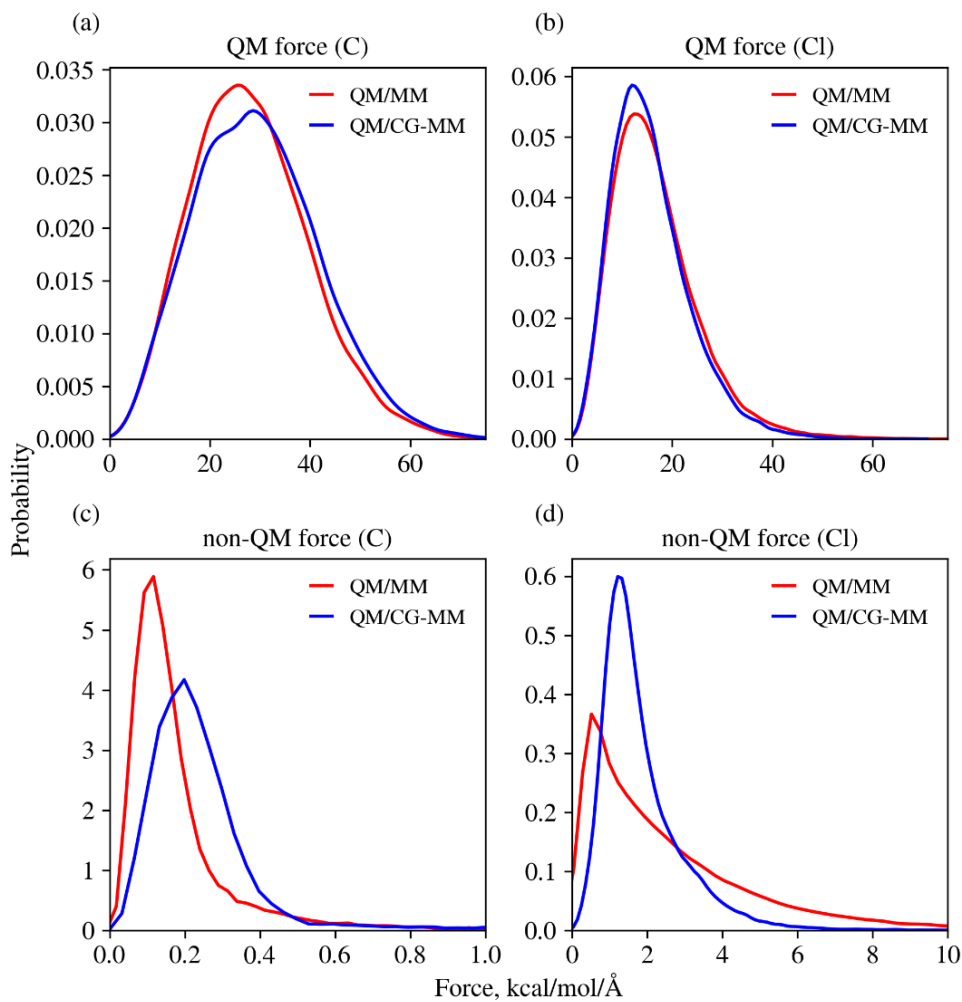

Figure 4. QM (a and b) and non-QM (c and d) force distributions on C (a and c) and Cl (b and d) atoms, as obtained from QM/MM (red) or QM/CG-MM (blue) calculations. The DFT/DZVP level of theory is used for the QM part.

*Radial distribution functions (RDF)*. In Figure 5 we compare AA and CG RDFs for C-CG, Cl-CG, and CG-CG pairs, computed with QM/MM or QM/CG-MM using both DFT/DZVP and DFTB potentials from Figure 3. Remarkably, the QM/CG-MM RDFs are nearly identical to those of QM/MM, apart from slight overstructuring. Moreover, DFT-QM/CG-MM with DFTB potentials also yields the liquid structure in very close agreement to QM/MM, with slight overstructuring in the solvation shell of Cl atoms, despite qualitative differences in potential shapes and large uncertainties (Figure 3). This observation suggests that RDFs between QM and CG particles are not too sensitive to fine details of CG potentials, and computationally less expensive DFTB-QM/MM calculations can be employed in the first step of Figure 1 to parameterize the CG field.

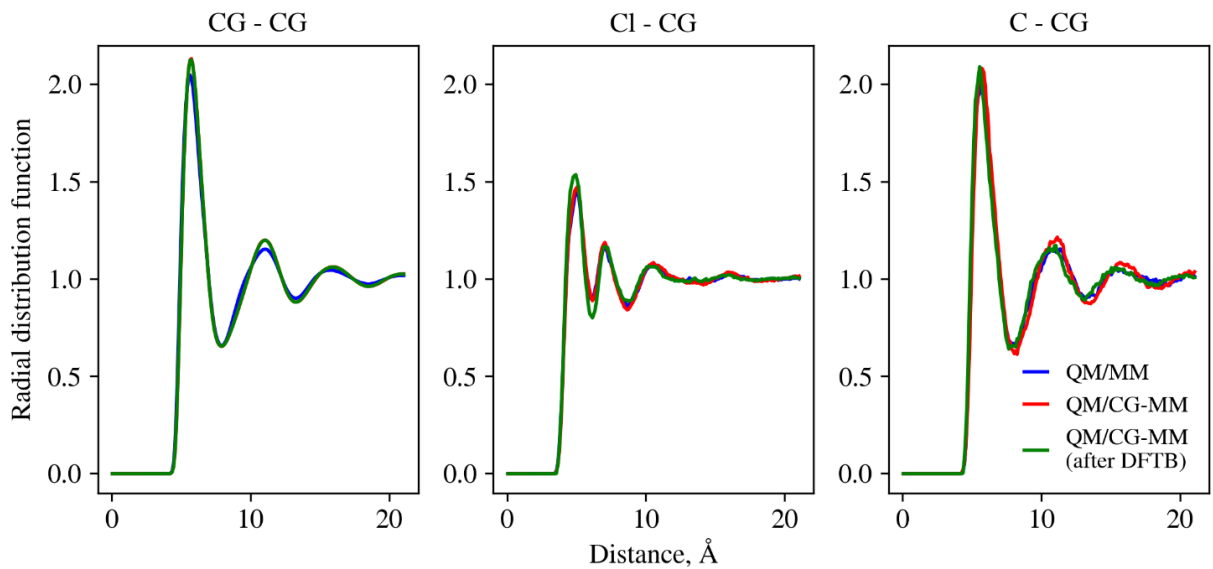

Figure 5. Radial distribution functions between CG particles, Cl and CG, and C and CG, computed using QM/MM (blue line), QM/CG-MM parameterized on DZVP QM/MM (red line), and QM/CG-MM parameterized on DFTB QM/MM (green line).

*Three-body distribution functions (3DF)*. To evaluate the performance of DFT-QM/CG-MM in the description of many-body effects, in Figure 6 we report distribution functions of CG-C-CG, CG-Cl-CG, and CG-CG-CG angles within the first solvation shell, weighted by $\sin\theta$ and normalized to unity. There is an excellent agreement between QM/CG-MM and QM/MM, which we attribute to the fundamental connection of the underlying force-matching MS-CG method to the Yvon-Born-Green liquid state theory, such that the critical three-body correlation effects are incorporated implicitly in pairwise CG potentials[52]. The first peak in all three angular distributions lies close to 60°, due to similar positions of first peaks in C-CG, Cl-CG, and CG-CG RDFs (5.65, 5.05, and 5.75 Å, respectively), such that all three particles form a slightly perturbed equilateral triangle at free energy minima. We attribute the bimodal distribution of CG-C-CG and CG-CG-CG to the spherical symmetry of the CG solvation shell around CG and an "averaged" C atom, whereas in the CG-Cl-CG distribution the bonded $CCl_3$ group occupies part of the shell, leading to monomodality.

*Source of excellent QM/CG-MM performance*. The high similarity of RDFs in Figure 5 from DFT- and DFTB-parameterized QM/CG-MM simulations is striking, particularly in light of Henderson's uniqueness theorem[81] stating that to a particular RDF there corresponds a unique effective pair potential. To explain this apparent contradiction, we recall that C and Cl atoms exhibit correlated motion due to the presence of C-Cl bonds, and thus it is the collective effect of C-CG and Cl-CG interactions that is physically significant, not the individual contributions. To show this more formally, we again separate the motion of C and Cl atoms into a translational, rotational, and vibrational motions. Then, we recognize that, since in the homogeneous liquid all directions are equivalent, and intramolecular vibrations are uncorrelated among particles due to the short time scale, vibrational and rotational contributions to statistical averages reduce to constants. Therefore, it is only the translational motion of entire $CCl_4$ molecules that is influenced by intermolecular interactions, and it is the sum of C-CG and Cl-CG interactions that influence the liquid structure. Therefore, if the negative deviations in the C-CG potential (when going from DFT to DFTB, for example) compensate for the positive deviations in the Cl-CG potential, or vice versa, the net intermolecular potential and thus the liquid structure should remain the same. Such compensation is already observed in Figure 3 if we imagine integrating negative potential branches – the integrated C-CG DFTB potential would be more negative than that of DFT, whereas the Cl-CG integral would correspondingly be more positive.

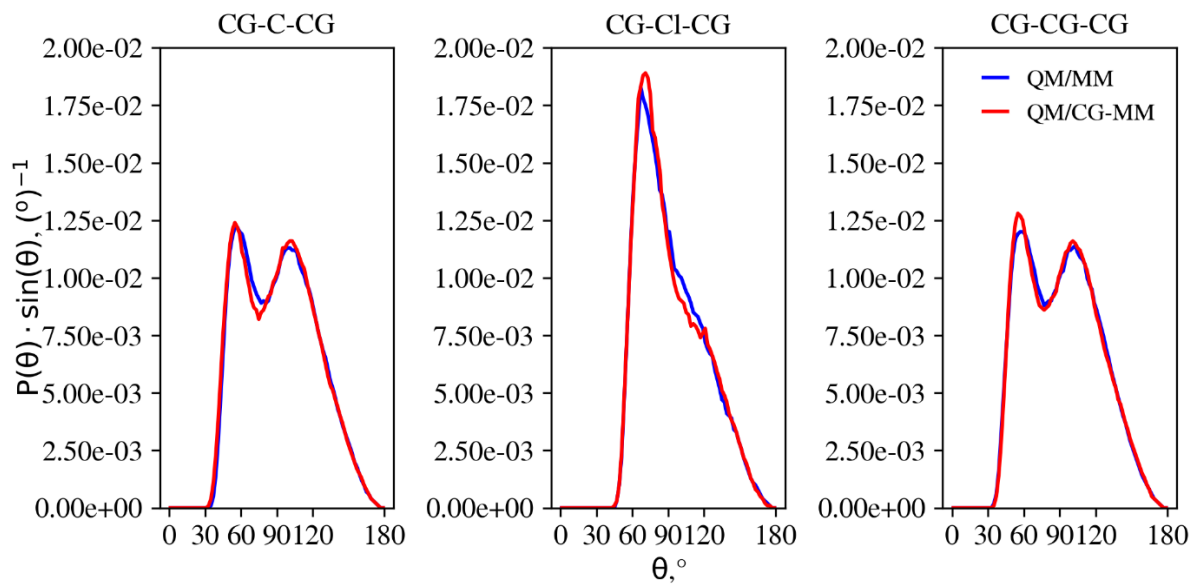

Figure 6. Three-body distribution functions for CG-C-CG, CG-Cl-CG, and CG-CG-CG angles, computed using QM/MM (blue) or QM/CG-MM (red). Binned angle values were multiplied by $sin\,\theta$ and normalized to unity.

To obtain more direct evidence to what we call the "compensation effect," in Figure 7 we report QM-CG force-matched potentials obtained by considering the entire QM region as a single bead, i.e., by coarse-graining C-CG and Cl-CG potentials into a single QM-CG interaction. Unlike the potentials in Figure 3, both DFT and DFTB potentials are remarkably similar, supporting the compensation hypothesis. Interestingly, the QM-CG potential also resembles the CG-CG potential despite the different fundamental treatment of underlying QM↔MM vs. MM↔MM electrostatic interactions. We attribute this to the fact that for uncharged species, van der Waals contributions, described through the same functional form in both types of interactions, dominate over electrostatics. Negligible electrostatic effects provide further support for the monopole approximation in eq. (11).

Figure 7 suggests that there are at least two ways of implementing QM/CG-MM. The first one is what we call the "1-to-1 QM/CG-MM", in which each QM atom type is mapped identically to a separate CG particle type (considered in this work). The second one is the "many-to-1 QM/CG-MM", in which several QM atoms are projected onto a fewer number of CG beads while retaining the AA resolution of the QM region and back mapping forces between beads onto QM atoms. Both methods have their advantages and drawbacks: the first approach is easier to implement, but produces noisy potentials and requires rather long trajectories for parameterization, while the second approach requires the implementation of force back mapping but yields more rapidly converging potentials, as follows from Figure 7.

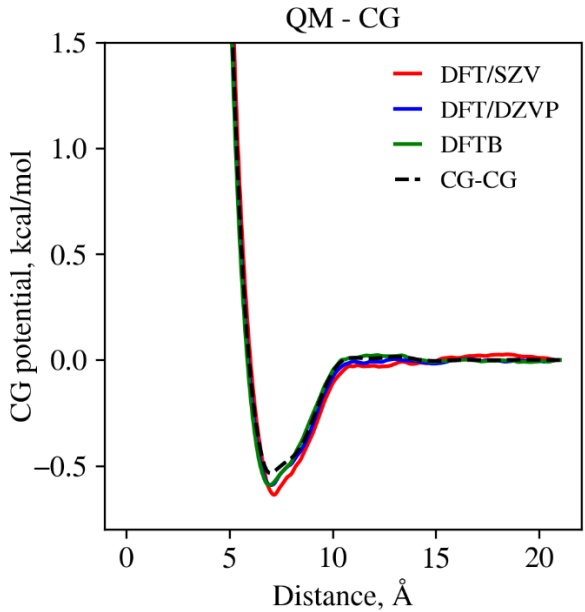

Figure 7. QM-CG potentials after force-matching, where the entire QM region is treated as a single CG bead. The CG-CG potential is shown for the reference.

We have made another interesting observation related to the CG potentials. It turns out that the potentials that yield accurate RDFs are rather far from being converged. In Figure 8 we compare the potentials computed from 1 ns vs. 2 ns DFT/DZVP-QM/MM trajectories. The CG-CG potential for the longer trajectory is ~ 0.15 kcal/mol deeper, whereas the C-CG potential is more negative by ~ 0.1 kcal/mol. More negative potentials for longer trajectories are associated with the fact that the system traverses a larger volume of the configuration space, which results in a more positive entropy. Since the CG potentials are effectively Landau free energies (potentials of the mean force) $F(r) = E(r) - TS(r)$, positive entropy $S(r)$ leads to more negative potentials, in agreement with Figure 8. A similar phenomenon has been observed by Sinitskiy and Voth [82] for lysozyme, who showed that the Taylor expansion coefficients of CG potentials are renormalized for longer simulation times $t$, varying as $t^{-\alpha_i}$ for $t$ up to 128 ns.

In order to show that the potential renormalization in our case is seemingly inconsequential for RDFs, it is instructive to consider the exact Yvon-Born-Green integral equation theory of liquids[83]. According to this theory, the quantity $kT \frac{\partial \ln g(r)}{\partial r}$, where $g(r)$ is the RDF, is an implicit functional of the potential derivative $\frac{dU(r)}{dr}$. It follows that the constant potential shift does not affect RDF. It turns out the 1 ns and 2 ns C-CG and CG-CG potentials primarily differ by the constant shift (-0.1 and -0.15 kcal/mol, respectively) in the interval of $0 < r < 10$ Å – shifted 1 ns potentials (dashed line in Figure 8) nearly coincide with the 2 ns potentials in this region. Beyond 10 Å the potentials are nearly zero, and thus the difference between them can be *approximated* by the square-well function with the threshold distance of ca. 10 Å. We conclude that RDFs from 1 ns and 2 ns trajectories should be nearly identical, apart from the contribution from the potential step at ~ 10 Å. Such a contribution is unlikely to have a significant influence on the RDF shape. $g(r) - 1$ is a sum of direct and indirect correlation functions[83], and the small details of the potential primarily influence the former (it becomes $e^{-U(r)/kT}$ at the low density limit). Direct correlation functions, however, decay rapidly and should be negligible by the second solvation shell peak at ~10 Å. Therefore, the not fully converged CG potential derived from the 1 ns trajectory yields an accurate RDF, as we observe in Figure 5.

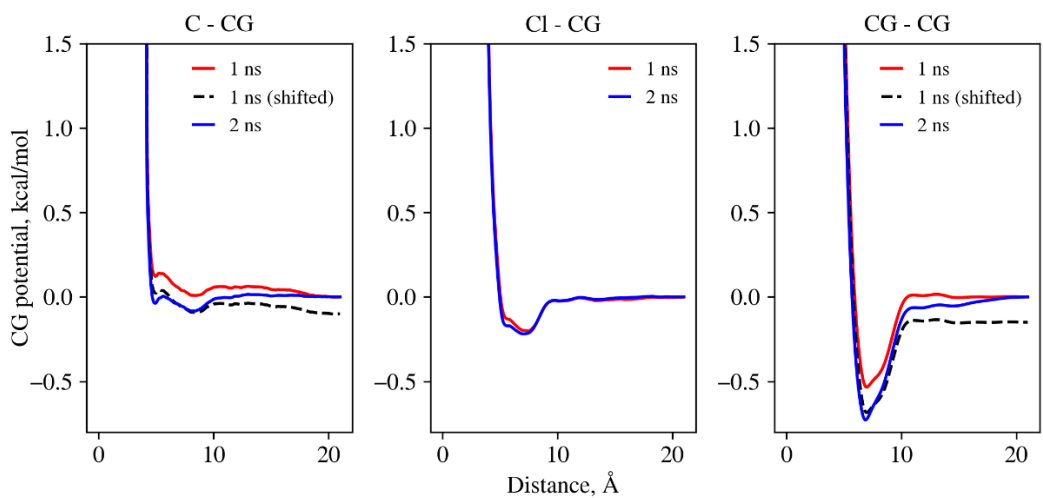

Figure 8. CG potentials determined by the force-matching procedure on DFT/DZVP-QM/MM trajectories of 1 ns (red) and 2 ns (blue) length. The dashed line corresponds to the 1 ns trajectory data, shifted by – 0.1 kcal/mol (C-CG) and – 0.15 kcal/mol (CG-CG).

**TBHC+benzyl radical reactive system.** Having demonstrated the promising performance of the DFT-QM/CG-MM method on a CCl$_4$ liquid, we will turn our attention to the tert-butyl hypochlorite (TBHC) reaction with the benzyl radical in CCl$_4$ (Figure 9). We compute the reaction free energy profile using a combination of the umbrella sampling and a weighted histogram analysis method, as described in the Computational Details section. We choose the reaction collective variable (CV) to be $r_{AC} - r_{AB}$, where A, B, C are chlorine, oxygen, and carbon atoms, respectively, and B and C belong to TBHC and benzyl (Figure 9). Since running long DFT-QM/MM MD is prohibitively expensive for this system, we take advantage of the fact that the liquid structure is insensitive to fine details of CG potentials (see Figure 5) due to the compensation effect and use the DFTB theory for the QM region to obtain CG potentials *only*. The total number of CG potentials to be parameterized equals 10, since $N_{QM,g} = 9$ and $N_{CG,g} = 1$ (see the Implementation Section). High-accuracy DFT-QM/CG-MM computations of the barrier using the computed CG potentials then follow.

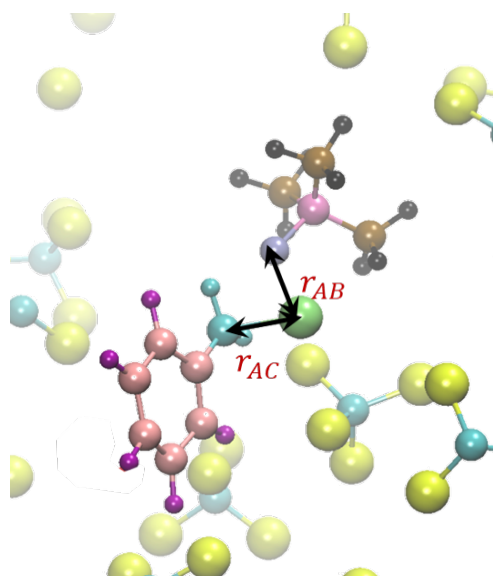

Figure 9. The reactive model system consisting of a benzyl radical and a tert-butoxy hypochlorite (TBHC) species. The reaction CV involves AB and AC interatomic distances between Cl (green), O (purple), and C(cyan) atoms.

In Figure 10 we report ten CG potentials from both biased and unbiased DFTB-QM/MM calculations. We find the potentials to be mainly attractive, except for those involving aromatic ($C_{ar}$) and sp² aliphatic ($C(CH_2)$) carbon atoms. On the opposite, their associated hydrogen atoms display particularly deep minima, hinting at the "compensation effect", described above. There is a significant dependence of CG potentials on the reaction coordinate, especially for $C(CH_2)$-CG, $H(CH_2)$-CG, Cl-CG, and O-CG interactions, which is unsurprising, given the bond topological changes during the reaction and associated CG population variations in first solvation shells. Part of observed variations are likely due to associated potential uncertainties, which we discussed above for the case of the CCl$_4$ liquid.

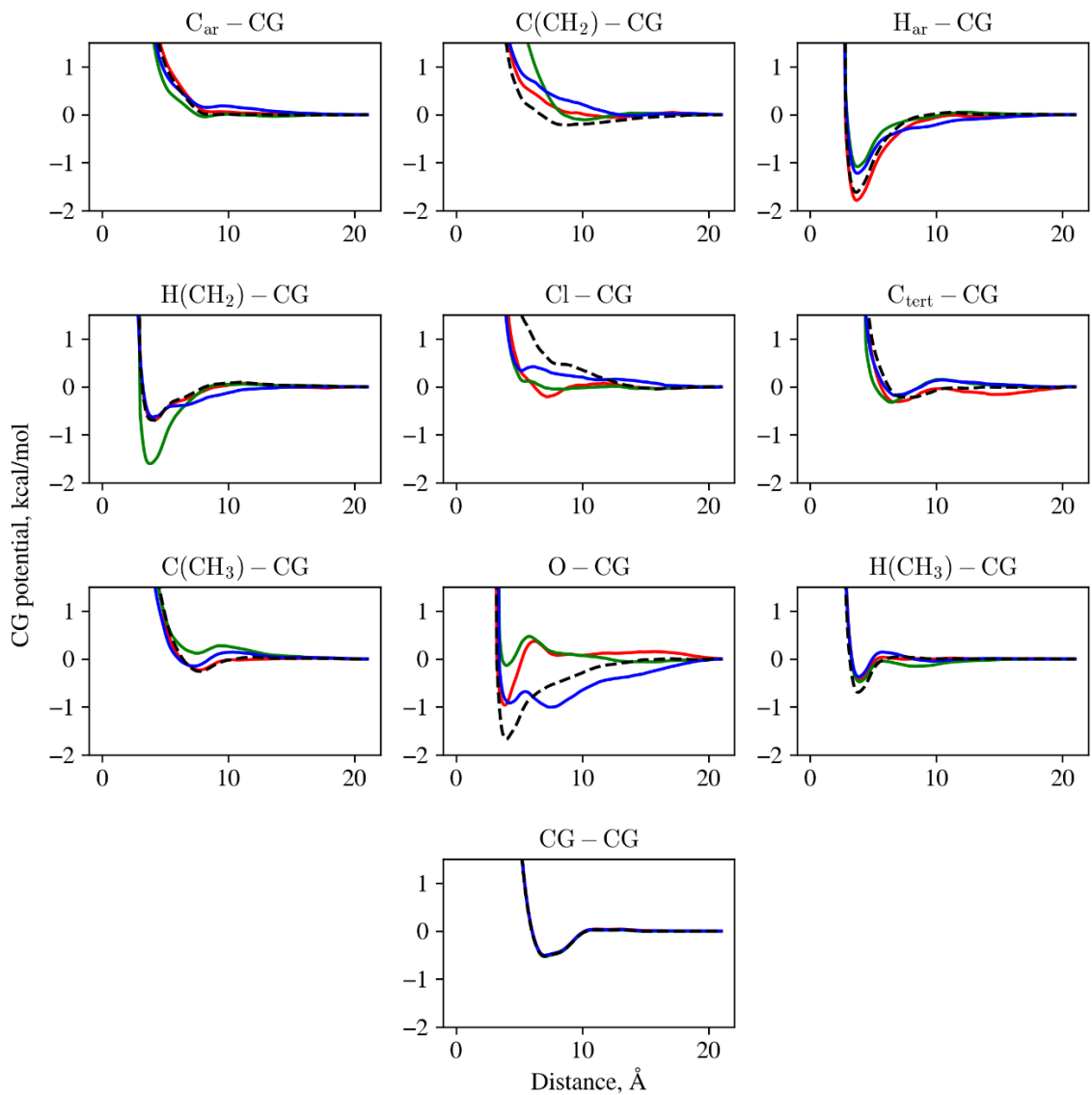

Figure 10. CG potentials describing interactions between CG $CCl_4$ particles and QM atoms in the TBHC/benzyl system at 313.15 K. Forces and trajectories from DFTB-QM/MM were used in parameterization. Curves correspond to an unbiased simulation (dashed) and biased ones with the target CV values of -4.75 Å (red), 0.0 Å (green), and +4.75 Å (blue).

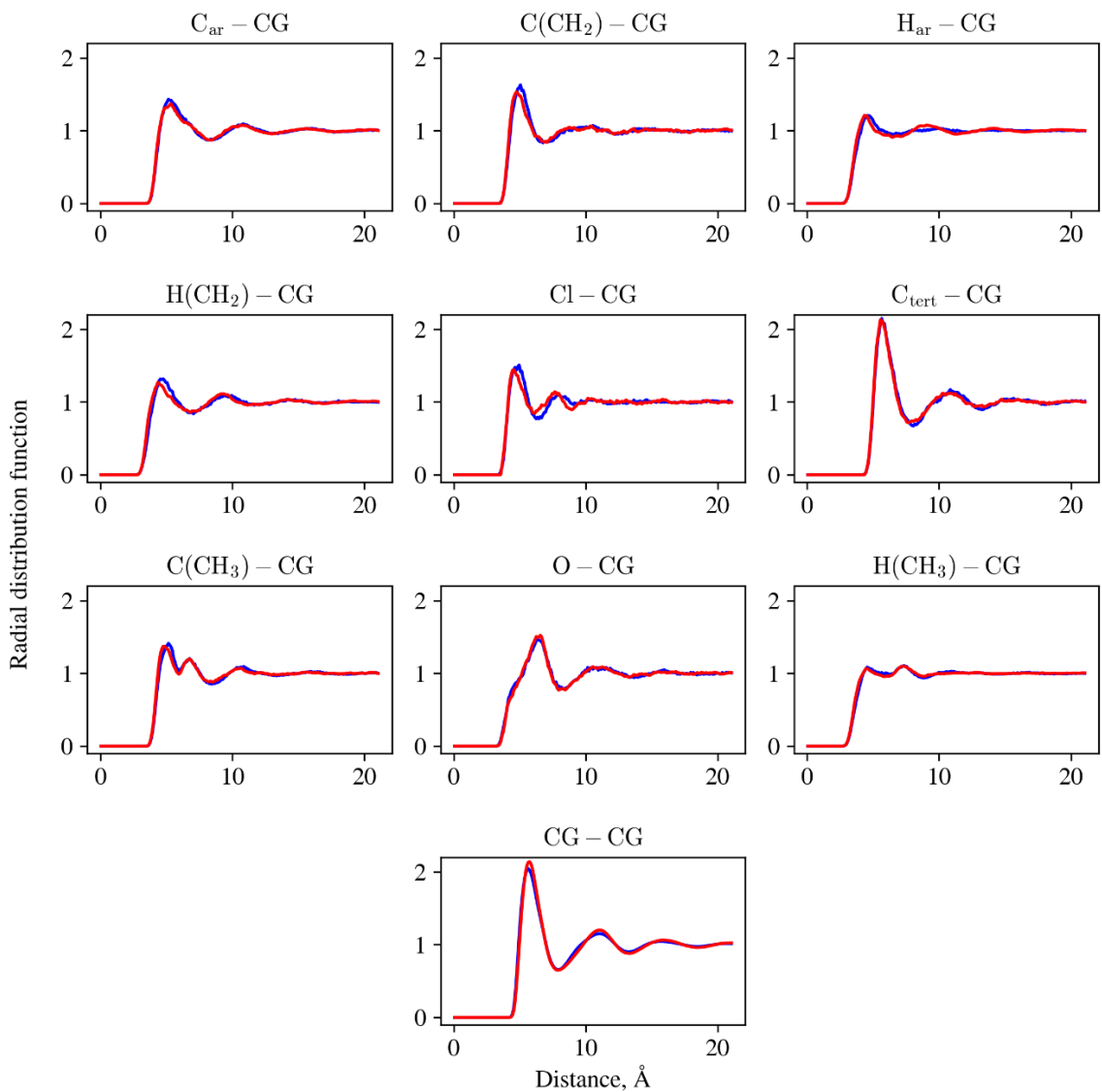

Figure 11. Radial distribution functions obtained in unbiased simulations of the TBHC+benzyl system in CCl$_4$ solvent using DFTB-QM/MM (blue) and DFTB-QM/CG-MM (red) approaches at 313.15 K.

In Figure 11 we compare DFTB-QM/MM and DFTB-QM/CG-MM radial distribution functions, obtained in unbiased simulations using unbiased CG potentials from Figure 10. The QM/CG-MM method reproduces all-atom spatial correlations very accurately. Unlike in simple liquids, deeper CG potentials in Figure 10 do not necessarily translate to sharper RDF peaks in Figure 11 – this is particularly seen for the H$_{ar}$-CG pairs. The origin of this phenomenon is two-fold: (1) other parts of the QM region occupy part of the solvation shell for a particular atom, affecting the RDF of CG beads; and (2) rapid correlated vibrations and rotations of atoms relative to centers of mass of individual functional groups (e.g. $C_6H_5$), independent of CG bead positions. The overall agreement between AA and CG RDFs indicates that the solvation environment structure around the reaction complex is described reliably by the DFTB-QM/CG-MM method.

Figure 12 compares reaction free energy profiles from umbrella sampling calculations for all-atom and coarse-grained resolution models. We employed the CG potentials from unbiased DFTB-QM/MM reference simulations. There is an excellent agreement between DFT-QM/MM and DFT-QM/CG-MM energy profiles near initial (IS) and final (FS) states – the coarse-graining method accurately reproduces the overall free energy of the reaction. At the peak, deviations up to 1-2 kcal/mol are observed, with slight statistically significant overstabilization of the transition state (TS) in the coarse-grained model. There are a number of various factors that may contribute to this error, including (1) the choice of the DFTB theory during the parameterization, (2) the neglect of electrostatic interactions, (3) CG basis set incompleteness, (4) neglected CG potential variations along the reaction coordinate, and (5) too short QM/MM MD trajectories. However, the $CCl_4$ liquid analysis, presented above, suggests that factors 1 and 2 are not significant. The RDF agreement rules out factor 3, and the near-perfect description of the IS-FS energy difference, despite qualitatively different CG potentials (see Figure 10), rules out factor 4. Finally, small error bars in Figure 12 make factor 5 unlikely.

To explain this TS overstabilization, we propose the following hypothesis. Upon the TS formation as AB+C→ [A-B-C], parts of first solvation shells of AB and C are replaced by C and AB fragments, respectively, with a simultaneous release of a few coordinated solvent molecules to the bulk. Consequently, the corresponding gain in the translational entropy of freed solvent molecules contributes to the free energy barrier, whereas the contributions due to internal $CCl_4$ degrees of freedom (DoFs) cancel out. In a two-state model, the change in the free energy due to this entropic gain may be expressed as $\Delta G = G_{bulk} - G_{shell}$, where $G_{bulk}$ and $G_{shell}$ are $CCl_4$ free energies in the bulk and the first solvation shell, respectively. Since AA and CG solvent representations yield similar RDFs, and RDFs are directly related to potentials of the mean force (PMF), $G_{shell}^{CG} \approx G_{shell}^{AA}$. However, $G_{bulk}^{CG} \neq G_{bulk}^{AA}$ for the following reason. Neglecting intramolecular DoFs, which do not contribute to the phenomenon under consideration, we note that, in contrast to the AA model, in which COM motion is stochastic after integrating out internal DoFs, in the MSCG model COMs undergo deterministic motion, leading to qualitative differences in dynamics and thermodynamics between two representations. In particular, the diffusivity $D$ is greater in the CG model[84]. Since $D$ is related to the excess entropy $\Delta s_{ex}$ as $D \sim \exp(\Delta s_{ex})$ through the Rosenfeld scaling relation[85], $\Delta s_{ex}$ is less negative in the CG model. However, $\Delta s_{ex} = S_{tr,bulk} - S_{tr,ideal}$, where $S_{tr,bulk}$ and $S_{tr,ideal}$ are translational contributions to bulk liquid and ideal gas entropies. Since $S_{ideal}^{CG} = S_{ideal}^{AA}$ and $\Delta s_{ex}^{CG} > \Delta s_{ex}^{AA}$, $S_{tr,bulk}^{CG} > S_{tr,bulk}^{AA}$ and $G_{bulk}^{CG} > G_{bulk}^{AA}$. This qualitative picture suggests that the release of the solvent translational entropy upon TS formation is greater in the CG representation, leading to the reaction barrier underestimation in the QM/CG-MM model, as observed in Figure 12.

We do not present a comparison with the experimental barrier here, as it is well-known that semilocal density functionals, such as PBE used in this work, underestimate transition state energies considerably due to the electron delocalization error[86]. Hybrid functionals that mix the semilocal XC energy and the exact exchange in various proportions demonstrate much better performance[86], but at the expense of a considerably higher computational cost. We note that the DFT-QM/CG-MM method is general by construction and can be used with any XC functional available in the software.

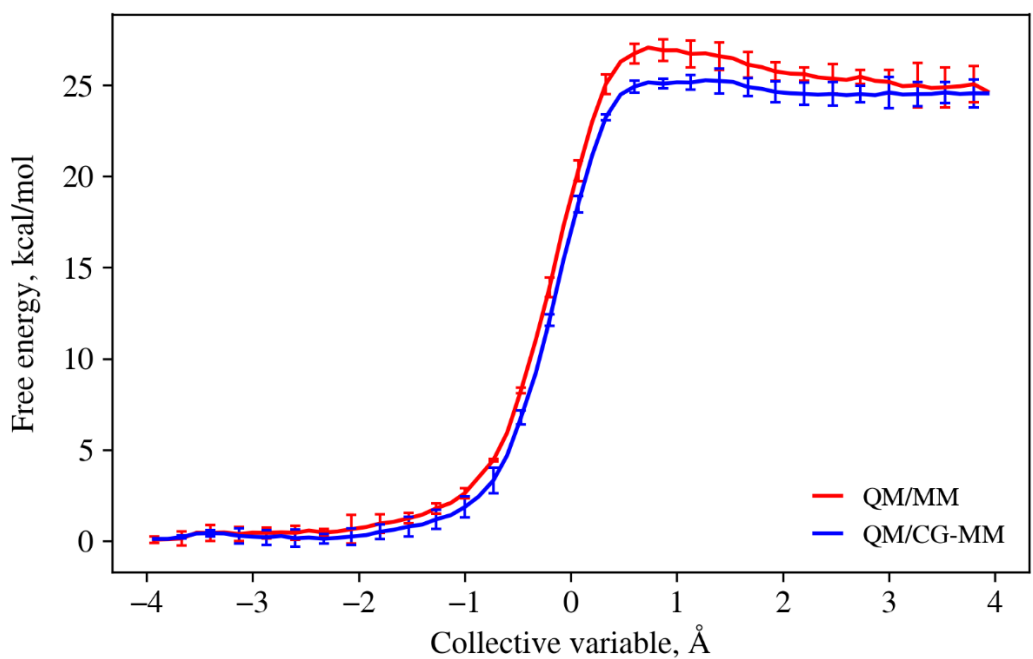

Figure 12. Free energy profile for the TBHC reaction with the benzyl radical along the reaction coordinate defined in Figure 8. Energies are computed with DZVP/DFT-QM/MM (red) and DZVP/DFT-QM/CG-MM for two initial geometry guesses of the reactive complex (blue and green) at 313.15 K. The QM/MM profile is averaged over three independent umbrella sampling/WHAM calculations. 95% confidence intervals are shown.

The computational gains achieved through the use of the DFT-QM/CG-MM method depend on the relative size of QM and MM regions. In the THBC+benzyl and "$CCl_4$-in-$CCl_4$" systems considered so far, the number of MM atoms is rather small (~2,500), and thus the computational advantage of coarse-graining is quite insignificant, as the evaluation of the QM energy dominates the cost. This is particularly evident in the first entry of Table 1: coarse-graining the MM part in the reactive system reduces the computational cost by a mere 7 %. To determine computational gains in the scenario of an extremely large MM region (as in a large solvated protein), we set up the "$CCl_4$-in-$CCl_4$" model system consisting of 5 QM and 67,495 MM atoms in a $129.54^3$ $Å^3$ simulation box. In this case, the QM/CG-MM method reduces the computational cost by 77% (second entry in Table 1). Expectedly, coarse-graining also reduces the equilibration time, as manifested in temperature equilibration requiring only 4 ps instead of 15 ps in QM/MM (Figure S2).

Table 1. Computational cost comparison between QM/MM and QM/CG-MM methods. Calculations were performed on 25 Intel E5-2680 CPU units.

| System | Elapsed time per MD time step, s | | Computational cost reduction, % |
| --- | --- | --- | --- |
| | DFT-QM/MM | DFT-QM/CG-MM | |
| THBC+benzyl (umbrella sampling; 29 QM+2,500 MM atoms) | 4.01* | 3.73* | 7 |
| $CCl_4$ QM in $CCl_4$ MM (5 QM+67,495 MM atoms) | 0.79 | 0.18 | 77 |

*Averaged over all umbrella sampling windows

## CONCLUSIONS

The systematic, "bottom-up" theory of Quantum Mechanics/Coarse-Grained Molecular Mechanics[48] has been expanded to the density functional theory in this work. The new DFT-QM/CG-MM formalism has been derived and implemented in the CP2K software and with the multiscale coarse-graining (MS-CG) force-matching technique[51]. We assessed the method's performance in comparison with AA Quantum Mechanics/Molecular Mechanics (QM/MM) using two model systems involving a non-polar solvent: $CCl_4$ liquid with a single molecule treated quantum mechanically and a reaction of tert-butyl hypochlorite and benzyl radical in $CCl_4$. We found that the method reproduces radial distribution functions of all-atom reference models, as well as the free energy surface along the reaction coordinate, quite accurately, with errors less than 1-2 kcal/mol. The method provides its significant computational speed-up from the MM coarse-graining in systems with large number ratios of MM to QM atoms, as might be expected and for which the method is intended. The generalization of the DFT-QM/CG-MM to treat polar systems will be presented in follow-up work.

## APPENDIX: DERIVATION OF ADDITIVE QM/MM WITH ELECTROSTATIC EMBEDDING FROM THE KOHN-SHAM THEORY

In the Kohn-Sham (KS) formalism[7] of DFT with semilocal exchange-correlation (XC) effects[8] and separate dispersion corrections[49,50], treated perturbatively, the total energy of a system of N atoms is defined as

$$E(\mathbf{R}^N) = \sum_i f_i \varepsilon_i - \int \rho(\mathbf{r}) \left(\frac{1}{2}\phi[\rho] + \mu_{xc,sl}(\rho(\mathbf{r}))\right) d\mathbf{r} + E_{xc,sl}[\rho] + E_{vdW}(\mathbf{R}^N) \\ + E_{NN}(\mathbf{R}^N), \quad (13)$$

where $\rho$ is the electron density, $\varepsilon_i$ are the energies of KS one-electron states, $f_i$ are their electron occupancies, $\phi$ is the Hartree potential, $E_{xc,sl}$ is the semilocal XC energy, $\mu_{xc,sl} = \partial E_{xc,sl}/\partial \rho$ is the semilocal XC potential, $E_{NN}$ is the internuclear electrostatic repulsion energy, $E_{vdW}$ is the energy of van der Waals interactions, and $\mathbf{R}^N$ are the positions of all nuclei. Energies $\varepsilon_i$ and their corresponding states $\psi_i$ are solutions to the KS equations,

$$H_{ks}[\rho; \mathbf{R}^N]|\psi_i\rangle = \varepsilon_i|\psi_i\rangle,$$

$$H_{ks} = -\frac{1}{2}\nabla^2 + v_N(\mathbf{r}; \mathbf{R}^N) + \phi[\mathbf{r}; \rho] + \mu_{xc,sl}(\rho(\mathbf{r}))$$

(14)

where $H_{ks}$ is the Kohn-Sham Hamiltonian operator, $-\frac{1}{2}\nabla^2$ is the kinetic energy operator (atomic units are used), and $v_N$ is the external (nuclear) potential. The electron density is defined as $\rho(\mathbf{r}) = \sum_i f_i|\psi_i(\mathbf{r})|^2$, and the states $|\psi_i\rangle$ are assumed to be expanded in the unspecified atomic basis set.

In the following, we divide the system into two non-overlapping parts I and II with nuclear coordinates $\mathbf{R}^{N_I}$ and $\mathbf{R}^{N_{II}}$ and electron densities $\rho_I$ and $\rho_{II}$, respectively, and assume that basis set functions defining two regions are orthogonal between regions, but are not orthogonal within each region. This division can be natural, if the boundary coincides with definitions of molecules, or artificial if the boundary cuts through a chemical bond. In the subsequent derivations, we consider the former case and treat nonorthogonality effects, which induce Pauli repulsion between molecules, perturbatively[87], assuming that the overlap between regions I and II is small. The KS Hamiltonian in eq. (14) then becomes

$$H_{ks} = -\frac{1}{2}\nabla^2 + v_N(\mathbf{r}; \mathbf{R}^{N_I}) + v_N(\mathbf{r}; \mathbf{R}^{N_{II}}) + \phi[\mathbf{r}; \rho_I] + \phi[\mathbf{r}; \rho_{II}] + \mu_{xc,sl}(\rho_I(\mathbf{r})) \\ + \mu_{xc,sl}(\rho_{II}(\mathbf{r}))$$

(15)

For non-overlapping densities, the corresponding KS equations $H_{ks}|\psi_i^I\rangle = \varepsilon_i^I|\psi_i^I\rangle$ and $H_{ks}|\psi_i^{II}\rangle = \varepsilon_i^{II}|\psi_i^{II}\rangle$ hold for each subsystem separately and $\varepsilon_i = \varepsilon_i^I + \varepsilon_i^{II}$, where $|\psi_i^k\rangle$ and $\varepsilon_i^k$ are the KS eigenstates and eigenvalues of the k$^{th}$ subsystem, respectively. This stems from the fact that, if KS states are expanded in localized basis set functions $|\varphi_a\rangle$ as $|\psi_i\rangle = \sum_{a\in I} c_a^i|\varphi_a\rangle + \sum_{c\in II} c_c^i|\varphi_c\rangle$, KS energies for the entire system are $\varepsilon_i = \sum_{a,b\in I} c_a^i c_b^i\langle\varphi_a|H_{ks}|\varphi_b\rangle + \sum_{c,d\in I} c_c^i c_d^i\langle\varphi_c|H_{ks}|\varphi_d\rangle + \sum_{a\in I,c\in II} c_a^i c_c^i\langle\varphi_a|H_{ks}|\varphi_c\rangle$. Groups of terms correspond to $\varepsilon_i^I$, $\varepsilon_i^{II}$, and the subsystems' coupling energy, respectively. However, the latter is zero, since $\langle\varphi_a|H_{ks}|\varphi_c\rangle \to 0$ for $a \in I$, $c \in II$, due to $\langle\varphi_a|\varphi_c\rangle \to 0$ for non-overlapping atoms, proving the independence of subsystems.

In the following, we identify the subsystem I with the QM region requiring full quantum-mechanical description, and the subsystem II with the MM region described with approximate energy expressions. We neglect the term $\mu_{xc,sl}(\rho_{MM}(\mathbf{r}))$ in $H_{ks}|\psi_i^{QM}\rangle = \varepsilon_i^{QM}|\psi_i^{QM}\rangle$ (see eq. (15)), since $\rho_{QM}(\mathbf{r}) = \sum_i f_i^{QM}|\psi_i^{QM}(\mathbf{r})|^2$ and $\rho_{MM}(\mathbf{r})$ are assumed to not overlap. Furthermore, we neglect the polarization of the MM electron density by the environment and approximate $v_N(\mathbf{r}; \mathbf{R}^{N_{II}}) + \phi[\mathbf{r}; \rho_{II}]$ by the electrostatic potential of fixed point charges at positions of atomic nuclei $\mathbf{R_i}$, as $V_{es}^{MM}(\mathbf{r}; \mathbf{R}^{N_{MM}}) = \sum_{i \in MM} q_{MM,i}/|\mathbf{r} - \mathbf{R_i}|$. This leads us to the following KS equation for the QM subsystem:

$$\left(-\frac{1}{2}\nabla^2 + v_N(\mathbf{r}; \mathbf{R}^{N_{QM}}) + \phi[\mathbf{r}; \rho_{QM}] + \mu_{xc,sl}\left(\rho_{QM}(\mathbf{r})\right) \right.$$
$$\left. + V_{es}^{MM}(\mathbf{r}; \mathbf{R}^{N_{MM}}| \mathbf{q}_{MM})\right) \psi_i^{QM}(\mathbf{r}) = \varepsilon_i^{QM} \psi_i^{QM}(\mathbf{r}), \qquad (16)$$

where $\mathbf{q}_{MM}$ is the set of MM charges.

To obtain the total energy expression, we note that the QM subsystem is typically small, and the MM subsystem is large, and thus the effect of the former on the latter can be neglected. Also, since chemical bonds in the MM subsystem remain intact, the molecular mechanics (MM) force field $E_{MM \leftrightarrow MM}(\mathbf{R}^{N_{MM}})$ provides a reasonable approximation to its energy as a function of atomic coordinates. Taking eq. (14) and (13) together, we obtain the equation that forms the basis of the electrostatic embedding QM/MM scheme,

$$E_{QM/MM}(\mathbf{R}^N) = E_{QM}(\mathbf{R}^{N_{QM}}; \rho_{QM}) + E_{MM \leftrightarrow MM}(\mathbf{R}^{N_{MM}}|\mathbf{q}_{MM}, \mathbf{K}_{MM})$$
$$+ E_{vdW,QM \leftrightarrow MM}(\mathbf{R}^N|\mathbf{K}_{vdW}) + E_{NN,QM \leftrightarrow MM}(\mathbf{R}^N|\mathbf{q}_{MM}) \qquad (17)$$
$$+ E_{ortho,QM \leftrightarrow MM}(\mathbf{R}^N|\mathbf{K}_{ortho}),$$

where the energy of the QM system is

$$E_{QM}(\mathbf{R}^{N_{QM}}; \rho_{QM}) =$$
$$\sum_i f_i^{QM} \varepsilon_i^{QM} - \int \rho_{QM}(\mathbf{r}) \left(\frac{1}{2}\phi[\mathbf{r}; \rho_{QM}] + \mu_{xc,sl}\left(\rho_{QM}(\mathbf{r})\right)\right) d\mathbf{r} + E_{xc,sl}[\rho_{QM}] \qquad (18)$$
$$+ E_{vdW}(\mathbf{R}^{N_{QM}}) + E_{NN}(\mathbf{R}^{N_{QM}}).$$

$E_{vdW,QM \leftrightarrow MM}$, $E_{NN,QM \leftrightarrow MM}$, and $E_{ortho,QM \leftrightarrow MM}$ account for long-range van der Waals, nuclear-nuclear interactions, and short-range nonorthogonality corrections (*vide supra*), respectively, between subsystems QM and MM.

We convert eq. (17) into a more convenient form by adding and subtracting the double-counting correction term $E_{dc,QM/MM} = -\int \rho_{QM}(\mathbf{r}) V_{es}^{MM}(\mathbf{r}; \mathbf{R}^{N_{MM}}| \mathbf{q}_{MM}) d\mathbf{r}$, yielding

$$E_{QM/MM}(\mathbf{R}^N) = E'_{QM}(\mathbf{R}^N; \rho_{QM}) + E_{MM \leftrightarrow MM}(\mathbf{R}^{N_{MM}}) + E_{QM \leftrightarrow MM}(\mathbf{R}^N) \qquad (19)$$

where $E'_{QM}(\mathbf{R}^N; \rho_{QM}) = E_{QM}(\mathbf{R}^{N_{QM}}; \rho_{QM}) + E_{dc,QM/MM}$ and $E_{QM \leftrightarrow MM}(\mathbf{R}^N) = E_{vdW,QM \leftrightarrow MM}(\mathbf{R}^N) + E_{ortho,QM \leftrightarrow MM}(\mathbf{R}^N) + E_{es,QM \leftrightarrow MM}(\mathbf{R}^N)$, and the electrostatic energy of QM/MM interaction is $E_{es,QM \leftrightarrow MM}(\mathbf{R}^N) = E_{NN,QM \leftrightarrow MM}(\mathbf{R}^N) + \int \rho_{QM}(\mathbf{r}) V_{es}^{MM}(\mathbf{r}; \mathbf{R}^{N_{MM}}| \mathbf{q}_{MM}) d\mathbf{r}$. The double-counting term cancels the corresponding contribution to eigenvalues $\varepsilon_i^{QM}$, so that dispersion, orthogonalization, and electrostatic interaction energies of QM and MM subsystems are combined in $E_{QM \leftrightarrow MM}$, whereas the polarization energy of the QM subsystem due to MM is included in $E'_{QM}$ only implicitly.

ACKNOWLEDGEMENTS

This research was supported by the Office of Naval Research, grant number N000141812574, and in part by the Kadanoff-Rice fellowship to A.M. from The University of Chicago Materials Research Science and Engineering Center, which is funded by the National Science Foundation under award number DMR-1420709. We thank Drs. Sriramvignesh Mani, Harshwardhan Katkar, Fikret Aydin, and Alex Pak for valuable discussions. We acknowledge the University of Chicago Research Computing Center for computational support of this work.

REFERENCES

(1) Knight, C.; Voth, G. A. The Curious Case of the Hydrated Proton. *Acc. Chem. Res.* **2012**, *45* (1), 101–109.

(2) Agmon, N.; Bakker, H. J.; Campen, R. K.; Henchman, R. H.; Pohl, P.; Roke, S.; Thämer, M.; Hassanali, A. Protons and Hydroxide Ions in Aqueous Systems. *Chem. Rev.* **2016**, *116* (13), 7642–7672.

(3) Wraight, C. A. Chance and Design—Proton Transfer in Water, Channels and Bioenergetic Proteins. *Biochim. Biophys. Acta (BBA)-Bioenergetics* **2006**, *1757* (8), 886–912.

(4) Voth, G. A. Computer Simulation of Proton Solvation and Transport in Aqueous and Biomolecular Systems. *Acc. Chem. Res.* **2006**, *39* (2), 143–150.

(5) Jain, A.; Ong, S. P.; Hautier, G.; Chen, W.; Richards, W. D.; Dacek, S.; Cholia, S.; Gunter, D.; Skinner, D.; Ceder, G. Commentary: The Materials Project: A Materials Genome Approach to Accelerating Materials Innovation. *Apl Mater.* **2013**, *1* (1), 11002.

(6) Hohenberg, P.; Kohn, W. Inhomogeneous Electron Gas. *Phys. Rev.* **1964**, *136* (3B), B864.

(7) Kohn, W.; Sham, L. J. Self-Consistent Equations Including Exchange and Correlation Effects. *Phys. Rev.* **1965**, *140* (4A), A1133.

(8) Perdew, J. P.; Burke, K.; Ernzerhof, M. Generalized Gradient Approximation Made Simple. *Phys. Rev. Lett.* **1996**, *77* (18), 3865.

(9) Čížek, J. On the Use of the Cluster Expansion and the Technique of Diagrams in Calculations of Correlation Effects in Atoms and Molecules. *Adv. Chem. Phys.* **1969**, 35–89.

(10) Raghavachari, K.; Trucks, G. W.; Pople, J. A.; Head-Gordon, M. A Fifth-Order Perturbation Comparison of Electron Correlation Theories. *Chem. Phys. Lett.* **1989**, *157* (6), 479–483.

(11) Fermi, E.; Pasta, P.; Ulam, S.; Tsingou, M. *Studies of the Nonlinear Problems*; Los Alamos

Scientific Lab., N. Mex., 1955.

(12) Verlet, L. Computer" Experiments" on Classical Fluids. I. Thermodynamical Properties of Lennard-Jones Molecules. *Phys. Rev.* **1967**, *159* (1), 98.

(13) McDonald, I. R.; Singer, K. Machine Calculation of Thermodynamic Properties of a Simple Fluid at Supercritical Temperatures. *J. Chem. Phys.* **1967**, *47* (11), 4766–4772.

(14) Valleau, J. P.; Card, D. N. Monte Carlo Estimation of the Free Energy by Multistage Sampling. *J. Chem. Phys.* **1972**, *57* (12), 5457–5462.

(15) Torrie, G. M.; Valleau, J. P. Monte Carlo Free Energy Estimates Using Non-Boltzmann Sampling: Application to the Sub-Critical Lennard-Jones Fluid. *Chem. Phys. Lett.* **1974**, *28* (4), 578–581.

(16) Kumar, S.; Rosenberg, J. M.; Bouzida, D.; Swendsen, R. H.; Kollman, P. A. The Weighted Histogram Analysis Method for Free-energy Calculations on Biomolecules. I. The Method. *J. Comput. Chem.* **1992**, *13* (8), 1011–1021.

(17) Roux, B. The Calculation of the Potential of Mean Force Using Computer Simulations. *Comput. Phys. Commun.* **1995**, *91* (1–3), 275–282.

(18) Valsson, O.; Tiwary, P.; Parrinello, M. Enhancing Important Fluctuations: Rare Events and Metadynamics from a Conceptual Viewpoint. *Annu. Rev. Phys. Chem.* **2016**, *67*, 159–184.

(19) Broyden, C. G. The Convergence of a Class of Double-Rank Minimization Algorithms 1. General Considerations. *IMA J. Appl. Math.* **1970**, *6* (1), 76–90.

(20) Mills, G.; Jónsson, H. Quantum and Thermal Effects in H 2 Dissociative Adsorption: Evaluation of Free Energy Barriers in Multidimensional Quantum Systems. *Phys. Rev. Lett.* **1994**, *72* (7), 1124.

(21) Henkelman, G.; Jónsson, H. Improved Tangent Estimate in the Nudged Elastic Band Method for Finding Minimum Energy Paths and Saddle Points. *J. Chem. Phys.* **2000**, *113* (22), 9978–9985.

(22) Henkelman, G.; Uberuaga, B. P.; Jónsson, H. A Climbing Image Nudged Elastic Band Method for Finding Saddle Points and Minimum Energy Paths. *J. Chem. Phys.* **2000**, *113* (22), 9901–9904.

(23) Eyring, H. The Activated Complex in Chemical Reactions. *J. Chem. Phys.* **1935**, *3* (2), 107–115.

(24) Warshel, A. *Computer Modeling of Chemical Reactions in Enzymes and Solutions*; John Wiley & Sons: New York, 1991.

(25) Wang, X.; Hu, J.; Liu, G.; Tian, J.; Wang, H.; Gong, M.; Liu, S. Reversibly Switching Bilayer Permeability and Release Modules of Photochromic Polymersomes Stabilized by Cooperative Noncovalent Interactions. *J. Am. Chem. Soc.* **2015**, *137* (48), 15262–15275.

(26) Ansar, S. M.; Kitchens, C. L. Impact of Gold Nanoparticle Stabilizing Ligands on the Colloidal Catalytic Reduction of 4-Nitrophenol. *ACS Catal.* **2016**, *6* (8), 5553–5560.

(27) Kohn, W. Density Functional and Density Matrix Method Scaling Linearly with the Number of Atoms. *Phys. Rev. Lett.* **1996**, *76* (17), 3168.

(28) Prodan, E.; Kohn, W. Nearsightedness of Electronic Matter. *Proc. Natl. Acad. Sci.* **2005**, *102* (33), 11635–11638.

(29) Westheimer, F. H. Steric Effects in Organic Chemistry. *by MS Newman, John Wiley Sons, New York* **1956**, 523–555.

(30) Lifson, S.; Warshel, A. Consistent Force Field for Calculations of Conformations, Vibrational Spectra, and Enthalpies of Cycloalkane and N-alkane Molecules. *J. Chem. Phys.* **1968**, *49* (11), 5116–5129.

(31) MacKerell Jr, A. D.; Bashford, D.; Bellott, M.; Dunbrack Jr, R. L.; Evanseck, J. D.; Field, M. J.; Fischer, S.; Gao, J.; Guo, H.; Ha, S. All-Atom Empirical Potential for Molecular Modeling and Dynamics Studies of Proteins. *J. Phys. Chem. B* **1998**, *102* (18), 3586–3616.

(32) Jorgensen, W. L.; Maxwell, D. S.; Tirado-Rives, J. Development and Testing of the OPLS All-Atom Force Field on Conformational Energetics and Properties of Organic Liquids. *J. Am. Chem. Soc.* **1996**, *118* (45), 11225–11236.

(33) Cornell, W. D.; Cieplak, P.; Bayly, C. I.; Gould, I. R.; Merz, K. M.; Ferguson, D. M.; Spellmeyer, D. C.; Fox, T.; Caldwell, J. W.; Kollman, P. A. A Second Generation Force Field for the Simulation of Proteins, Nucleic Acids, and Organic Molecules. *J. Am. Chem. Soc.* **1995**, *117* (19), 5179–5197.

(34) Warshel, A.; Levitt, M. Theoretical Studies of Enzymic Reactions: Dielectric, Electrostatic and Steric Stabilization Of. *J. mol. Biol* **1976**, *103*, 227–249.

(35) Friesner, R. A.; Guallar, V. Ab Initio Quantum Chemical and Mixed Quantum Mechanics/Molecular Mechanics (QM/MM) Methods for Studying Enzymatic Catalysis. *Annu. Rev. Phys. Chem.* **2005**, *56*, 389–427.

(36) Senn, H. M.; Thiel, W. QM/MM Methods for Biomolecular Systems. *Angew. Chemie Int. Ed.* **2009**, *48* (7), 1198–1229.

(37) Shaik, S.; Cohen, S.; Wang, Y.; Chen, H.; Kumar, D.; Thiel, W. P450 Enzymes: Their Structure, Reactivity, and Selectivity   Modeled by QM/MM Calculations. *Chem. Rev.* **2010**, *110* (2), 949–1017.

(38) Benediktsson, B.; Bjornsson, R. QM/MM Study of the Nitrogenase MoFe Protein Resting State: Broken-Symmetry States, Protonation States, and QM Region Convergence in the FeMoco Active Site. *Inorg. Chem.* **2017**, *56* (21), 13417–13429.

(39) Laino, T.; Mohamed, F.; Laio, A.; Parrinello, M. An Efficient Real Space Multigrid QM/MM Electrostatic Coupling. *J. Chem. Theory Comput.* **2005**, *1* (6), 1176–1184.

(40) Landau, L. D.; Lifshitz, E. M. Statistical Physics, Part 1: Volume 5. **1980**.

(41) Sun, R.; Sode, O.; Dama, J. F.; Voth, G. A. Simulating Protein Mediated Hydrolysis of ATP and Other Nucleoside Triphosphates by Combining QM/MM Molecular Dynamics with Advances in Metadynamics. *J. Chem. Theory Comput.* **2017**, *13* (5), 2332–2341.

(42) Katkar, H. H.; Davtyan, A.; Durumeric, A. E. P.; Hocky, G. M.; Schramm, A. C.; Enrique, M.; Voth, G. A. Insights into the Cooperative Nature of ATP Hydrolysis in Actin Filaments. *Biophys. J.* **2018**, *115* (8), 1589–1602.

(43) Noid, W. G. Perspective: Coarse-Grained Models for Biomolecular Systems. *J. Chem. Phys.* **2013**, *139* (9), 09B201_1.

(44) Saunders, M. G.; Voth, G. A. Coarse-Graining Methods for Computational Biology. *Annu. Rev. Biophys.* **2013**, *42*, 73–93.

(45) Brini, E.; Algaer, E. A.; Ganguly, P.; Li, C.; Rodriguez-Ropero, F.; van der Vegt, N. F. A. Systematic Coarse-Graining Methods for Soft Matter Simulations–a Review. *Soft Matter* **2013**, *9* (7), 2108–2119.

(46) Ingólfsson, H. I.; Lopez, C. A.; Uusitalo, J. J.; de Jong, D. H.; Gopal, S. M.; Periole, X.; Marrink, S. J. The Power of Coarse Graining in Biomolecular Simulations. *Wiley Interdiscip. Rev. Comput. Mol. Sci.* **2014**, *4* (3), 225–248.

(47) Dannenhoffer-Lafage, T.; Voth, G. A. Reactive Coarse-Grained Molecular Dynamics. *J. Chem. Theory Comput.* **2020**.

(48) Sinitskiy, A. V; Voth, G. A. Quantum Mechanics/Coarse-Grained Molecular Mechanics (QM/CG-MM). *J. Chem. Phys.* **2018**, *148* (1), 14102.

(49) Tkatchenko, A.; Scheffler, M. Accurate Molecular van Der Waals Interactions from Ground-State Electron Density and Free-Atom Reference Data. *Phys. Rev. Lett.* **2009**, *102* (7), 73005.

(50) Grimme, S.; Antony, J.; Ehrlich, S.; Krieg, H. A Consistent and Accurate Ab Initio Parametrization of Density Functional Dispersion Correction (DFT-D) for the 94 Elements H-Pu. *J. Chem. Phys.* **2010**, *132* (15), 154104.

(51) Izvekov, S.; Voth, G. A. Multiscale Coarse Graining of Liquid-State Systems. *J. Chem. Phys.* **2005**, *123* (13), 134105.

(52) Noid, W. G.; Chu, J.-W.; Ayton, G. S.; Voth, G. A. Multiscale Coarse-Graining and Structural Correlations: Connections to Liquid-State Theory. *J. Phys. Chem. B* **2007**, *111* (16), 4116–4127.

(53) Noid, W. G.; Chu, J.-W.; Ayton, G. S.; Krishna, V.; Izvekov, S.; Voth, G. A.; Das, A.; Andersen, H. C. The Multiscale Coarse-Graining Method. I. A Rigorous Bridge between Atomistic and Coarse-Grained Models. *J. Chem. Phys.* **2008**, *128* (24), 244114.

(54) Izvekov, S.; Voth, G. A. A Multiscale Coarse-Graining Method for Biomolecular Systems. *J. Phys. Chem. B* **2005**, *109* (7), 2469–2473.

(55) Shell, M. S. The Relative Entropy Is Fundamental to Multiscale and Inverse Thermodynamic Problems. *J. Chem. Phys.* **2008**, *129* (14), 144108.

(56) Hammer, B.; Nørskov, J. K. Theory of Adsorption and Surface Reactions. In *Chemisorption and Reactivity on Supported Clusters and Thin Films*; Springer, 1997; pp 285–351.

(57) Hutter, J.; Iannuzzi, M.; Schiffmann, F.; VandeVondele, J. Cp2k: Atomistic Simulations of Condensed Matter Systems. *Wiley Interdiscip. Rev. Comput. Mol. Sci.* **2014**, *4* (1), 15–25.

(58) VandeVondele, J.; Krack, M.; Mohamed, F.; Parrinello, M.; Chassaing, T.; Hutter, J. Quickstep: Fast and Accurate Density Functional Calculations Using a Mixed Gaussian and Plane Waves Approach. *Comput. Phys. Commun.* **2005**, *167* (2), 103–128.

(59) Pople, J. A.; Beveridge, D. L. Approximate Molecular Orbital Theory, 1970. McGraw-Hill, New York.

(60) Lu, L.; Izvekov, S.; Das, A.; Andersen, H. C.; Voth, G. A. Efficient, Regularized, and Scalable Algorithms for Multiscale Coarse-Graining. *J. Chem. Theory Comput.* **2010**, *6* (3), 954–965.

(61) Jewett, A. I.; Zhuang, Z.; Shea, J.-E. Moltemplate a Coarse-Grained Model Assembly Tool. *Biophys. J.* **2013**, *104* (2), 169a.

(62) Martínez, L.; Andrade, R.; Birgin, E. G.; Martínez, J. M. PACKMOL: A Package for Building Initial Configurations for Molecular Dynamics Simulations. *J. Comput. Chem.* **2009**, *30* (13), 2157–2164.

(63) Humphrey, W.; Dalke, A.; Schulten, K. VMD: Visual Molecular Dynamics. *J. Mol. Graph.* **1996**, *14* (1), 33–38.

(64) Marrink, S. J.; Risselada, H. J.; Yefimov, S.; Tieleman, D. P.; De Vries, A. H. The MARTINI Force Field: Coarse Grained Model for Biomolecular Simulations. *J. Phys. Chem. B* **2007**, *111* (27), 7812–7824.

(65) Plimpton, S. *Fast Parallel Algorithms for Short-Range Molecular Dynamics*; Sandia National Labs., Albuquerque, NM (United States), 1993.

(66) Zavitsas, A. A.; Blank John, D. Kinetics of the Free-Radical Chain Chlorination of Hydrocarbons by Tert-Butyl Hypochlorite. *J. Am. Chem. Soc.* **1972**, *94* (13), 4603–4608.

(67) Nosé, S. A Unified Formulation of the Constant Temperature Molecular Dynamics

Methods. *J. Chem. Phys.* **1984**, *81* (1), 511–519.

(68) Hoover, W. G. Canonical Dynamics: Equilibrium Phase-Space Distributions. *Phys. Rev. A* **1985**, *31* (3), 1695.

(69) Shinoda, W.; Shiga, M.; Mikami, M. Rapid Estimation of Elastic Constants by Molecular Dynamics Simulation under Constant Stress. *Phys. Rev. B* **2004**, *69* (13), 134103.

(70) LIPPERT, B. G.; PARRINELLO, J. H. and M. A Hybrid Gaussian and Plane Wave Density Functional Scheme. *Mol. Phys.* **1997**, *92* (3), 477–488.

(71) Lippert, G.; Hutter, J.; Ballone, P.; Parrinello, M. Response Function Basis Sets: Application to Density Functional Calculations. *J. Phys. Chem.* **1996**, *100* (15), 6231–6235.

(72) Goedecker, S.; Teter, M.; Hutter, J. Separable Dual-Space Gaussian Pseudopotentials. *Phys. Rev. B* **1996**, *54* (3), 1703.

(73) VandeVondele, J.; Hutter, J. An Efficient Orbital Transformation Method for Electronic Structure Calculations. *J. Chem. Phys.* **2003**, *118* (10), 4365–4369.

(74) Essmann, U.; Perera, L.; Berkowitz, M. L.; Darden, T.; Lee, H.; Pedersen, L. G. A Smooth Particle Mesh Ewald Method. *J. Chem. Phys.* **1995**, *103* (19), 8577–8593.

(75) Cordero, B.; Gómez, V.; Platero-Prats, A. E.; Revés, M.; Echeverría, J.; Cremades, E.; Barragán, F.; Alvarez, S. Covalent Radii Revisited. *Dalt. Trans.* **2008**, No. 21, 2832–2838.

(76) Laino, T.; Mohamed, F.; Laio, A.; Parrinello, M. An Efficient Linear-Scaling Electrostatic Coupling for Treating Periodic Boundary Conditions in QM/MM Simulations. *J. Chem. Theory Comput.* **2006**, *2* (5), 1370–1378.

(77) Elstner, M.; Porezag, D.; Jungnickel, G.; Elsner, J.; Haugk, M.; Frauenheim, T.; Suhai, S.; Seifert, G. Self-Consistent-Charge Density-Functional Tight-Binding Method for Simulations of Complex Materials Properties. *Phys. Rev. B* **1998**, *58* (11), 7260.

(78) Tribello, G. A.; Bonomi, M.; Branduardi, D.; Camilloni, C.; Bussi, G. PLUMED 2: New Feathers for an Old Bird. *Comput. Phys. Commun.* **2014**, *185* (2), 604–613.

(79) Grossfield, A. An Implementation of WHAM: The Weighted Histogram Analysis Method, Version 2.0.9. *Available Membr. urmc. rochester. edu/content/wham. Accessed June* **2014**, *1*, 2017.

(80) Cui, Q.; Elstner, M.; Kaxiras, E.; Frauenheim, T.; Karplus, M. A QM/MM Implementation of the Self-Consistent Charge Density Functional Tight Binding (SCC-DFTB) Method. *J. Phys. Chem. B* **2001**, *105* (2), 569–585.

(81) Henderson, R. L. A Uniqueness Theorem for Fluid Pair Correlation Functions. *Phys. Lett. A* **1974**, *49* (3), 197–198.

(82) Sinitskiy, A. V; Voth, G. A. Coarse-Graining of Proteins Based on Elastic Network Models. *Chem. Phys.* **2013**, *422*, 165–174.

(83) McQuarrie, D. A. Statistical Thermodynamics. **1973**.

(84) Izvekov, S.; Voth, G. A. Modeling Real Dynamics in the Coarse-Grained Representation of Condensed Phase Systems. American Institute of Physics 2006.

(85) Rosenfeld, Y. Relation between the Transport Coefficients and the Internal Entropy of Simple Systems. *Phys. Rev. A* **1977**, *15* (6), 2545.

(86) Galano, A.; Alvarez-Idaboy, J. R. Kinetics of Radical-molecule Reactions in Aqueous Solution: A Benchmark Study of the Performance of Density Functional Methods. *J. Comput. Chem.* **2014**, *35* (28), 2019–2026.

(87) Albright, T. A.; Burdett, J. K.; Whangbo, M.-H. *Orbital Interactions in Chemistry*; John Wiley & Sons, 2013.

**For Table of Contents Only**

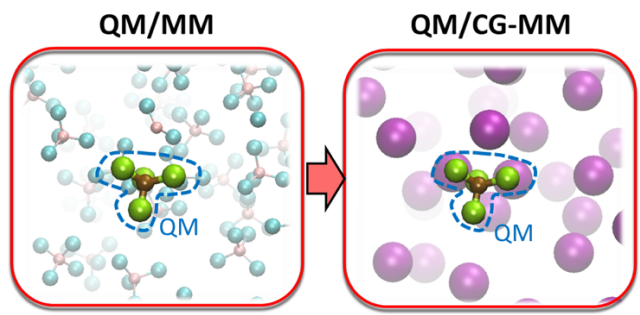

**Supporting Information for**

**Density Functional Theory-based Quantum Mechanics/Coarse-Grained Molecular Mechanics: Theory and Implementation**

Alexander V. Mironenko and Gregory A. Voth

*Department of Chemistry, Chicago Center for Theoretical Chemistry, James Franck Institute, and Institute for Biophysical Dynamics,*
The University of Chicago, Chicago, Illinois 60637

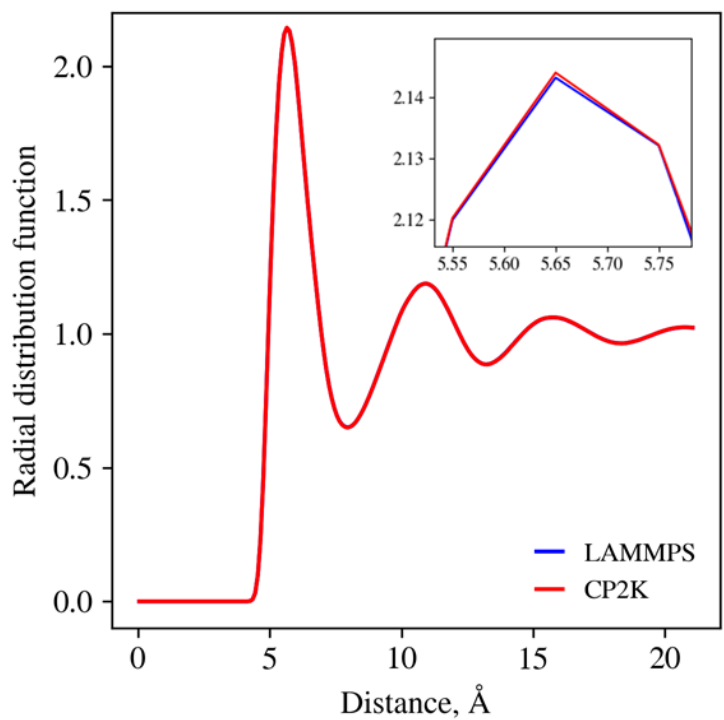

Figure S1. Comparison of radial distribution functions for the CG $CCl_4$ model, as obtained using LAMMPS (blue) or CP2K with the newly implemented tabulated potential (red).

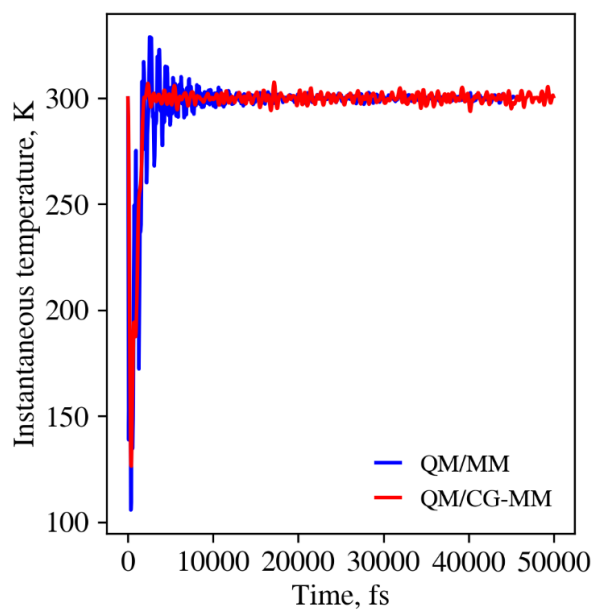

Figure S2. Instantaneous temperature evolution during the equilibration period of QM/MM (blue) vs QM/CG-MM (red) MD simulation of the "$CCl_4$ QM in $CCl_4$ MM" system with 67,495 MM atoms.